\documentclass[reprint, superscriptaddress, amsmath,amssymb, aps]{revtex4-2}

\usepackage{amsmath}
\usepackage{amsfonts}
\usepackage{amssymb}
\usepackage{graphicx}
\usepackage{bm}

\usepackage[normalem]{ulem}

\begin{document}

\title{Stability and decay of subradiant patterns in a quantum gas with photon-mediated interactions} 

\author{Alexander Baumgärtner}
\thanks{these authors contributed equally}
\affiliation{Institute for Quantum Electronics, Eidgen\"ossische Technische Hochschule Z\"urich, Otto-Stern-Weg 1, CH-8093 Zurich, Switzerland}
\affiliation{Current address: JILA, University of Colorado and National Institute of Standards and Technology, and Department of Physics, University of Colorado, Boulder, Colorado 80309, USA}

\author{Simon Hertlein}
\thanks{these authors contributed equally}
\affiliation{Institute for Quantum Electronics, Eidgen\"ossische Technische Hochschule Z\"urich, Otto-Stern-Weg 1, CH-8093 Zurich, Switzerland}

\author{Tom Schmit}
\thanks{these authors contributed equally}
\affiliation{Theoretical Physics, Saarland University, Campus E2.6, D-66123 Saarbr\"ucken, Germany}

\author{Davide Dreon}
\affiliation{Institute for Quantum Electronics, Eidgen\"ossische Technische Hochschule Z\"urich, Otto-Stern-Weg 1, CH-8093 Zurich, Switzerland}
\affiliation{Current address: planqc, Lichtenbergstr. 8, 85748 Garching b. München, Germany}

\author{Carlos  M\'{a}ximo}
\affiliation{Institute for Quantum Electronics, Eidgen\"ossische Technische Hochschule Z\"urich, Otto-Stern-Weg 1, CH-8093 Zurich, Switzerland}
\affiliation{Current address: Departamento de F\'{i}sica, Universidade Federal de S\~{a}o Carlos, 13565-905 S\~{a}o Carlos, S\~{a}o Paulo, Brazil}

\author{Xiangliang Li}
\affiliation{Institute for Quantum Electronics, Eidgen\"ossische Technische Hochschule Z\"urich, Otto-Stern-Weg 1, CH-8093 Zurich, Switzerland}
\affiliation{Current address: Beijing Academy of Quantum Information Sciences, Xi-bei-wang East Road No.10, 100193 Beijing, China}

\author{Giovanna Morigi}
\affiliation{Theoretical Physics, Saarland University, Campus E2.6, D-66123 Saarbr\"ucken, Germany}

\author{Tobias Donner}\email{donner@phys.ethz.ch}
\affiliation{Institute for Quantum Electronics, Eidgen\"ossische Technische Hochschule Z\"urich, Otto-Stern-Weg 1, CH-8093 Zurich, Switzerland}

\date{\today}

\begin{abstract} 
Metastability and its relaxation mechanisms challenge our understanding of the stability of quantum many-body systems, revealing a gap between the microscopic dynamics of the individual components and the effective descriptions used for macroscopic observables. We observe excited self-ordered subradiant patterns in a quantum gas coupled to two optical cavities and report lifetimes far beyond the system's typical timescales. These patterns eventually decay through an abrupt transition reordering the atoms into a superradiant phase. Ab-initio theory fully captures this macroscopic behavior, revealing that the subradiant patterns are stabilized by photon-mediated long-range interactions, thereby manifesting universal features of metastability characteristic of long-range interacting systems, as in astrophysics and plasma physics. Our work sheds light on the microscopic mechanisms stabilizing quantum states of matter and highlights the potential of photon-mediated forces for engineering correlations in many-body quantum systems.
\end{abstract}

\maketitle

\section*{Introduction.}
Collective phenomena are ubiquitous in nature, influencing systems as diverse as material properties, economic markets, and galaxy formation. They typically emerge far from equilibrium, where metastability and relaxation dynamics play a crucial role in state transitions. Manifestations are the emergence of organized patterns, critical dynamics, and temporal phases in complex systems. Their understanding and control motivates the need to identify universal principles that underlie these behaviors. In the atomic realm, subradiance and superradiance offer striking examples of how collective effects manifest themselves in light-scattering ensembles, demonstrating the impact of synchronized and desynchronized interactions at microscopic scales~\cite{Dicke1954Coherence,Gross1982Superradiance,Brandes2005Coherent,Reitz2022Cooperative}. Subradiance occurs when emitters, in an antisymmetric superposition, oscillate out of phase and decouple from their electromagnetic environment, suppressing emission rates~\cite{Crubellier1985Superradiance,Devoe1996Observation,Barnes2005Far,Filip2011Preparation,McGuyer2015Precise,Reimann2015Cavity,verde2024spinselective,Sonnefraud2010Experimental}. Conversely, superradiance arises from symmetric superpositions, enhancing emission through constructive interference~\cite{Gross1982Superradiance,Ferioli2023Non}. 

Enhancement or suppression of light scattering is controlled by the relative phase among the emitters, which is typically determined by their spatial arrangement~\cite{Fernandez:2007,Neuzner:2016}. In atomic and molecular gases, superradiant states can self-enhance as the mechanical forces of the scattered light pulls emitters into phase-coherent patterns~\cite{Inouye1999Superradiant}. In optical cavities, light scattering creates effective long-range forces that even can stabilize ordered configurations of ultra-cold atomic gases~\cite{Ritsch:2013,Mivehvar2021Cavity}. One prominent example is the transition from a disordered superfluid to a self-ordered supersolid in a quantum gas coupled to a mode of the electromagnetic field of a high-finesse cavity, which is paradigmatic of the Dicke quantum phase transition~\cite{Hepp1973On,Wang1973Phase,Baumann2010Dicke,Garraway2011Dicke,Nataf2010No-go}. However, such self-organization mechanisms are generally inaccessible for subradiant ensembles, as scattered light -- and consequently, optical forces -- vanishes. For this reason, so far only transient subradiant states have been observed in itinerant systems~\cite{Guerin2016Subradiance,Cipris2021Subradiance,Ferioli2021Storage,Glicenstein2022From,Cola2009Recoil,Wolf2018Observation}, while stable subradiant patterns have been achieved by pinning the atoms using external potentials~\cite{Neuzner:2016,Yan2023Superradiant,Rui2020Subradiant}.

In this work, we report the measurement of a self-stabilized subradiant many-body phase. The phase is observed in a quantum gas interacting with two cavity fields, imposing competing order, as illustrated in Fig.\ \ref{fig:concept}\textbf{A}, \textbf{B}. The subradiant pattern is stabilized by the mechanical forces due to the coupling with a secondary cavity field, which pins the atoms in a pattern inhibiting emission into a primary cavity, as shown in Fig.\ \ref{fig:concept}\textbf{C}. Remarkably, the atoms remain locked into the subradiant pattern for exceptionally long times, even when the state is energetically unfavorable. The subradiant state eventually decays by means of a violent relaxation, at which point the gas abruptly reorganizes into a superradiant, stationary pattern. We find that atomic losses are solely responsible for the finite lifetime of the subradiant states, triggering a non-equilibrium, driven-dissipative phase transition from subradiance to superradiance.

These characteristics strikingly distinguish cavity-mediated subradiance from the exponential decay of subradiance in free space~\cite{Guerin2016Subradiance,Ferioli2021Storage,Rui2020Subradiant}, suggesting that global metastability and abrupt reorganization are intrinsically due to the long-range, cavity-mediated forces\cite{Campa:2009,Defenu:2023,Defenu:2024}.
We support this hypothesis by a microscopic theory that qualitatively and quantitatively aligns with the experimental results. This model allows us to show that the subradiant pattern is a fixed point of the collective dynamics. In the configuration landscape, the subradiant state is the local minimum of 
the selforganized potential and the number of atoms determines the height of the barrier, as illustrated in Fig.~\ref{fig:concept}\textbf{D}. Atom loss lowers the barrier until it reaches the stability threshold~\cite{Jain:2007,Campa:2008}, inducing the sudden decay from the subradiant to the superradiant pattern. This transition manifests characteristics of violent relaxation, common in instabilities of astrophysical and plasma clusters~\cite{Lynden-Bell:1968,Barre:2004,Campa:2008,Campa:2009}, underscoring the broader relevance of strong long-range interacting systems. Our theoretical framework fully captures the non-equilibrium phase transition from the subradiant into the superradiant pattern, shedding light onto the underlying microscopic processes.

\section*{Results}
\begin{figure}[]
\begin{center}
\includegraphics[width=1\columnwidth]{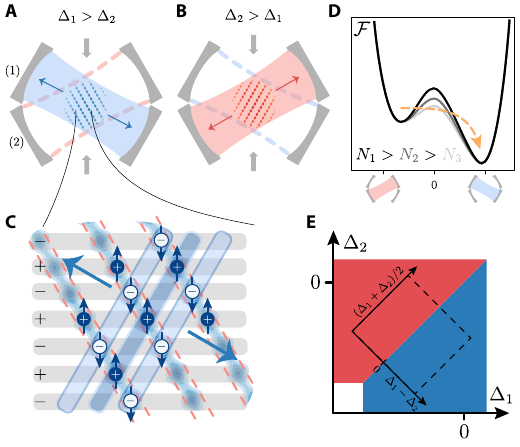}
\caption{\textbf{Subradiant state of a Bose-Einstein condensate in two crossed cavities.} 
\textbf{A}, \textbf{B} Self-ordered BEC (blue, red structures) simultaneously coupled to two optical cavities (1, 2) and transverse pump lattice (grey arrows). Depending on the relative detunings $\Delta_{1,2}$ between pump frequency and cavity resonances, the equilibrium structure scatters light either into cavity 1 or 2. 
\textbf{C} Microscopic illustration of the collective scattering. The grey lines indicate the standing wave pump field, its time phase is indicated by the ($\pm$) signs. The atoms (density pattern / spheres) oscillate with the driving field and order in  stripes that maximize emission into cavity 1, where all atoms within a plane (blue boxes) emit with the same phase. At the same time, they suppress scattering into cavity 2, where neighboring atoms emit in opposition of phase (red dashed planes).
\textbf{D} Free energy landscape illustrating the self-organized potential confining the subradiant phase. The barrier separates the subradiant from the superradiant phase, its height is determined by the atom number (see grey shading of lines). Atom losses lower the barrier and induce an abrupt transition into the superradiant order.
\textbf{E} Illustration of the stationary phase diagram (blue and red regions correspond to ordering in cavity 1 or 2, respectively) as a function of detunings $\Delta_{1,2}$ and for a fixed pump amplitude. The rotated axis and the dashed rectangle sketch the coordinate system used in the subsequent figures. The points where the bare cavity detunings vanish, $\Delta_{1,2}=0$, lie within the self-ordered phases since the dispersive coupling with the medium shifts the cavity resonances by several MHz.}
\label{fig:concept}
\end{center}
\end{figure}

\paragraph*{System description.}
The setup is composed of two high-finesse optical cavities ($\nu=$ 1, 2) that cross under an angle of 60$^\circ$, see Fig.\ ~\ref{fig:concept}\textbf{A}, \textbf{B}. A quantum gas of $^{87}$Rb atoms forms a BEC of $N=2.7(3)\cdot10^5$ and is placed in a dipole trap located at the crossing point of the modes. The cavities are characterized by their couplings $g_{\nu=1,2}/2\pi = (1.95(1), 1.77(1))$~MHz with the atomic dipolar transition, and the field decay rates $\kappa_{\nu=1,2}/2\pi = (147(4), 800(11))$~kHz. In the plane spanned by the cavities, the atoms are illuminated by a standing wave laser at frequency $\omega_p$ and with Rabi frequency $\Omega$. This pump field is far detuned by $\Delta_a = \omega_p - \omega_a = 2\pi \times 69.8(1)$~GHz on the blue with respect to the atomic resonance frequency $\omega_a$. It forms a repulsive optical lattice potential of depth $V_p= \Omega^2/\Delta_a$ that tends to localize the atoms at the nodes. 
The dispersive dynamics of the BEC-cavity system gives rise to the self-consistent formation of atomic density gratings that scatter light into the cavity and cause a shift in the cavity frequencies. The strength of these processes can be controlled by the detunings $\Delta_{\nu}=\omega_p - \omega^c_\nu$ between the driving laser field and the respective cavity resonance $\omega_{\nu}^c$~\cite{Morales2017Coupling}.

We do not expect cavity dissipation to play an important role in the observed dynamics. This might seem surprising, since in some protocols presented in this work the bare cavity detuning $\Delta_{\nu}$ vanishes. However, the effective cavity detunings are shifted by the collective dispersive shift to $\tilde{\Delta}_\nu\approx \Delta_\nu-N U_\nu/2$, where $U_\nu=g_\nu^2/\Delta_a>0$ is the dispersive Stark shift per particle due to the coupling with cavity mode $\nu=1,2$. In the experiment, $|\tilde{\Delta}_\nu|$ is of the order of several MHz and thus much larger than the cavity decay rates, even when the bare cavity detuning is zero, $\Delta_\nu=0$. The collective dispersive shift depends on the overlap between the atomic cloud and the cavity mode, and therefore it is a dynamical quantity. We thus present the data as a function of the bare cavity detuning $\Delta_\nu$.

\paragraph*{Stationary Phase Diagram.}
Fig.~\ref{fig:concept}\textbf{E} illustrates the stationary phase diagram as a function of the detunings $\Delta_\nu$ for a fixed pump strength $V_p$. The white region is the normal phase where the cavity field fluctuates about the vacuum and the atoms distribute along the minima of the standing wave pump field, parallel to the grey stripes of Fig.\ \ref{fig:concept}\textbf{C}. The blue and red regions correspond to stripe patterns coherently emitting into either cavity 1 or 2, respectively. The stripe patterns are sketched in subplots \textbf{A} and \textbf{B} extending to regions where the bare pump lattice potential is maximal. These configurations are facilitated by a finite cavity field: along the stripes, cavity and laser fields destructively interfere~\cite{Zupancic2019P},  and the resulting potential is reduced. This mechanism is the optomechanical analog of the dynamics of localized dipoles in cavities \cite{Zippilli_2004,Neuzner:2016}. The emerging patterns are such that simultaneous superradiant emission into both cavities is suppressed for sufficiently low pump strengths.

\begin{figure}[]
\begin{center}
\includegraphics[width=1\columnwidth]{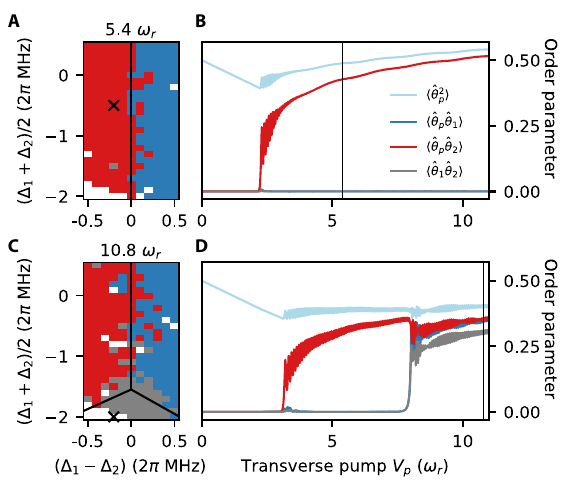}
\caption{\textbf{Steady state phase diagrams and occupation of atomic modes.}  \textbf{A}, \textbf{C} Experimental steady  state phase diagrams for pump lattice depths $V_p = (5.4, 10.8)~\omega_r$, with $\omega_\textrm{r}= 2\pi \times 3.77\,\mathrm{kHz}$. The blue (red) color indicates light exclusively in cavity 1 (2), white the absence of light, while the grey denotes coexistence, where atoms scatter in both cavities simultaneously. Black lines mark the phase boundaries calculated from the simulations of the GPE in \eqref{Eq:GPE}. A systematic error of $2\pi \times 0.1$~MHz in the cavity detunings has been taken into account~\cite{SM}. 
\textbf{B}, \textbf{D} GPE simulations of the expectation values of the order parameters $\langle \hat\theta_j\hat\theta_k\rangle$ with $j,k=p,1,2$ as a function of lattice depth $V_p$ for the detunings indicated by crosses in \textbf{A}, \textbf{C}, respectively. Vertical lines mark the lattice depths of panels \textbf{A, C}, respectively. Increasing $V_p$ initially leads to decreasing $\langle \hat{\theta}^2_p \rangle$, until self-ordering sets in and atoms are pulled towards the intensity maxima of the repulsive lattice potential. In panel \textbf{D} the two pump thresholds are visible: the first for self-ordering in cavity 2 and the second for self-ordering in both cavities. Note that the thresholds depend on the detunings $\Delta_1,\Delta_2$.}
\label{fig:coexistence}
\end{center}
\end{figure}

In order to experimentally determine the steady state phase diagram, we first fix the detunings $\Delta_1$, $\Delta_2$ and thus the expected crystalline configuration at a given lattice depth. We then linearly increase the pump lattice depth $V_p$ up to a target value~\cite{SM}, and record in real time the light leaking from the cavities using two heterodyne detection setups. We report the measured phase diagrams as a function of the relative cavity detuning $(\Delta_1-\Delta_2)$ and the mean cavity detuning $(\Delta_1+\Delta_2)/2$, which provide a convenient way to illustrate the protocols we implement in the next sections. The left panels of Fig.~\ref{fig:coexistence} show the measured phase diagrams for two target values of the pump amplitude.
We observe three regimes: For low pump lattice depths $V_p$, no self-ordering takes place and the cavities remain empty~\cite{SM}. For a large range of intermediate pump lattice depths, only the cavity closer to  resonance with the pump is filled with light, see Fig.~\ref{fig:coexistence}\textbf{A}).
Finally, for sufficiently deep pump lattices a region of coexistence becomes possible, where both cavities are populated simultaneously, see the grey region in Fig.~\ref{fig:coexistence}\textbf{C}. 

The experimental phase diagrams are reproduced by the ground state of a modified Gross-Pitaevskii equation (GPE), for the macroscopic wave function $\psi$ of the condensate in two dimensions,
\begin{equation}
\label{Eq:GPE}
    {\rm i}\hbar\partial_t\psi  = \Bigg(-\frac{\hbar^2\nabla^2}{2m} +\sum_{\nu=x,y}\frac{1}{2}m\omega_\nu^2 \nu^2  
    + V_0N|\psi|^2+\hat{H}_\mathrm{mf}[\psi] \Bigg) \psi\,,
\end{equation}
where $\hbar$ is the reduced Planck constant, $V_0$ is the $s$-wave, contact potential and $\hat{H}_\mathrm{mf}$ describes the optical potential in the mean-field approximation~\cite{SM}. For blue atomic detuning, $\Delta_a>0$, the optical potential can be cast into the form
\begin{eqnarray}
		\hat{H}_\mathrm{mf}[\psi]/\hbar &=&\left(\sqrt{V_p}\hat{\theta}_p + \sum_{\nu = 1}^2\sqrt{U_\nu} \alpha_\nu[\psi]\hat\theta_\nu\right)^2
\label{eq:mf_Hamiltonian:0}\,.
\end{eqnarray}
The variables $\alpha_\nu[\psi]$ are the electric field amplitudes of cavity mode $\nu$ and depend on the atomic density distribution. Finally, the phase-matching conditions are determined by operators $\hat \theta_j=\cos(\bm{k}_j \cdot \hat{\bm{x}})$ (with $j=p,1,2$) which capture the overlap between atomic density and the wave vector  $\bm{k}_j$ of the pump beam or the respective cavity. Equation \eqref{eq:mf_Hamiltonian:0} holds when cavity losses are negligible compared to the detuning of the dispersively shifted cavity resonance, i.e., $\kappa_{\nu} \ll |\Delta_\nu - U_\nu N\langle\hat{\theta}_\nu^2\rangle|=|\tilde{\Delta}_\nu|$. Our theoretical analysis incorporates the actual cavity dissipation rates. We have also benchmarked Eq.\ \eqref{Eq:GPE} against the full master equation, finding that dissipation has negligible impact within the explored parameter regime. Therefore, we will exclusively refer to Eq.\ \eqref{Eq:GPE} in the following.

Numerical analysis and experimental data show that there are two pump threshold values for self-organization: A lower one, separating the normal phase from the phase where the gas forms stripes emitting into one cavity mode, and a second, larger threshold, at which the atoms can form checkerboard patterns that simultaneously emit into both modes. 
The order parameters are shown in Fig.~\ref{fig:coexistence}\textbf{B}, \textbf{D} as a function of the pump strength $V_p$ for two different sets of detunings. The two thresholds are visible in Fig.~\ref{fig:coexistence}\textbf{D}. Here, simultaneous self-organization in both cavity modes is signaled by a finite value of the order parameter $\langle \hat{\theta}_1\hat{\theta}_2\rangle$. Such data would be experimentally accessible via destructive time-of-flight images~\cite{Morales2017Coupling,Zupancic2019P}. 

Insight into this behavior is gained from the minimization of the potential energy $E_\mathrm{mf}=\langle \hat{H}_\mathrm{mf}\rangle$. For blue detuning the potential energy is always positive, $E_\mathrm{mf}\ge 0 $, and minimization is trivially achieved when the atoms align along the nodes of the pump lattice, illustrated by the horizontal grey stripes in Fig.\ \ref{fig:concept}\textbf{C}. This configuration becomes mechanically unstable for increasing $V_p$: the new stable patterns then emit light into the cavities while minimizing the potential energy. This can occur through destructive interference between cavity fields and pump and can be summarized by the simplified   
condition $\sqrt{V_p} \ge -\left(\sqrt{U_1}\alpha_1 + \sqrt{U_2}\alpha_2\right)$ for which $E_{\rm mf}=0$ \cite{SM}. This condition constrains the amplitudes of the cavity fields, setting an upper bound to the values they can take. Assume, for instance, stripes at the wave vector $\bm{k}_p+\bm{k}_1$, such that $\alpha_1\neq 0$ and $\alpha_2=0$. The constrain on the cavity field amplitude permits to estimate the maximal depth of the potential confining the stripe, $\bar{V}_1\cos((\bm {k}_p+\bm{k}_1)\cdot\bm{x})$, such that it reads \cite{SM}
\begin{equation}
\label{eq:Vbar}
    \bar V_1 =  {V_p}\frac{U_1 N}{U_1 N/2 - \Delta_1}\,.
\end{equation}
Mechanical stability requires that the potential height is larger than the mechanical energy transferred to an atom by photon scattering, i.e., $|\bar{V}_\nu |>{\mathcal C}\omega_r$, where $\mathcal C\gtrsim1$ is a constant. Therefore, the stripe is (meta)stable when the pump exceeds the threshold value $V_p^{\mathrm{th}_\nu} = \mathcal{C}\omega_r (U_\nu N/2-\Delta_\nu)/(U_\nu N)$. This argument shows that self-organization of, say, cavity 1 is favored when $\bar V_1>\bar V_2$. In other words, the atoms self-organize in the mode with the largest dimensionless coupling strength $C_\nu=NU_\nu/\tilde{\Delta}_\nu$. Extending these arguments, we also estimate that the threshold for simultaneous self-organization in both cavity modes (phase coexistence) is $ V_p^{\mathrm{th}_{12}}\approx 2V_p^{\mathrm{th}_\nu}$ \cite{SM}, which qualitatively agrees with the numerical simulation in Fig.~\ref{fig:coexistence}\textbf{D}. The depth of the self-organized potential in Equation \eqref{eq:Vbar} depends nonlinearly on the number of atoms. 

In the rest of this work we will focus on the parameter regime of phase exclusion, where at equilibrium the two types of order compete (either $\langle \hat{\theta}_p \hat{\theta}_1 \rangle\neq 0$ or $\langle \hat{\theta}_p \hat{\theta}_2 \rangle\neq 0$, while $\langle \hat{\theta}_1\hat{\theta}_2\rangle=0$). In this regime, self-organization is expected to take place in the more strongly coupled cavity mode and the stationary phase is determined -- for constant pump intensity -- by the ratio $C_1/C_2$, and thus by the experimental parameters ($\Delta_1, \Delta_2$).

\begin{figure}[]
\begin{center}
\includegraphics[width=1\columnwidth]{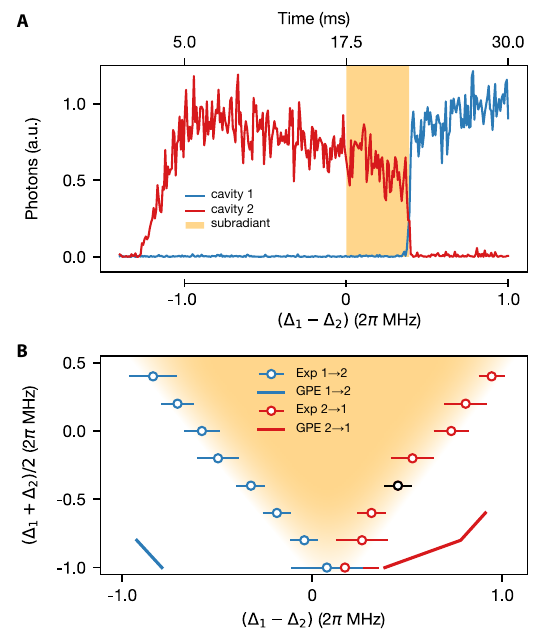}
\caption{\textbf{Hysteresis measurement.} \textbf{A} Real time evolution of the cavity occupations. In the first 5~ms, $V_p$ is ramped to 8.8~$\omega_r$ at fixed detunings. Afterwards, $V_p$ and the mean detuning $(\Delta_1+\Delta_2)/2$ are kept constant, while the relative detuning $(\Delta_1-\Delta_2)$ is ramped in time, passing the point $\Delta_1 = \Delta_2$ at 17.5~ms (see lower axis of this plot). The cavity populations only switch past this point at a positive relative detuning, indicating hysteretic behavior (orange color), where the BEC is in the subradiant pattern. 
\textbf{B} Switching points extracted from data sets as shown in panel \textbf{A} as a function of the mean detuning $(\Delta_1+\Delta_2)/2$. In the chosen parameter regime, (meta)stable co-existence does not occur. Errorbars display the standard deviation, the orange shading indicates the metastable region of subradiance observed experimentally. The black data point indicates the detuning at which the data in panel \textbf{A} was taken. The solid lines show results from GPE simulations, which predict a far broader region of hysteresis than observed in the experiment. For most mean detuning values, the GPE simulation does not predict any switch out of the subradiant state within the simulated experimental runtime.}
\label{fig:hysteresis}
\end{center}
\end{figure}

\paragraph*{Formation of Subradiant States.}
Having two competing superradiant crystals allows us to explore the transition between these structures \emph{in-situ}. In our protocol, we keep the pump lattice depth constant and vary the cavity mode resonance frequencies, and thus the detunings $\Delta_1$ and $\Delta_2$, as a function of time. We initialize the system in the stable superradiant state of one cavity mode, such as cavity 2, by setting \(\Delta_1 < \Delta_2\) (with $|\tilde{\Delta}_1|>|\tilde{\Delta}_2|$), by increasing the lattice depth $V_p$ to 8.8~$\omega_r$ within 5\,ms, and then keeping it constant. This lattice depth is above the self-organization threshold of cavity 2 but inside the regime of phase exclusion, so that the atoms form stripes emitting superradiantly into cavity 2 while cavity 1 is empty.  At \(t = 5\) ms, we linearly change the relative cavity detuning \((\Delta_1 - \Delta_2)\) over 25 ms to the inverse situation, \(\Delta_1 > \Delta_2\), while maintaining a constant mean detuning \((\Delta_1 + \Delta_2) / 2\). A typical trace of the cavity occupations is displayed in Fig.~\ref{fig:hysteresis}\textbf{A}. We record the signatures of self-organization in cavity 2 during the ramp. At 5~ms, the photon number in cavity 2 starts to decay since the effective coupling $C_2$ decreases during the sweep of the relative detuning. From the phase diagram (Fig.~\ref{fig:coexistence}\textbf{A}), one expects that the atoms should reorganize at $\Delta_1 \sim \Delta_2$. Instead, we observe a pronounced hysteretic behavior, indicated by the orange shading in Fig.~\ref{fig:hysteresis}\textbf{A}. Only past this point of balanced coupling, the system changes emission into cavity 1 and thereby displays an increase in the photon number, following the increase of $C_1$ over time.
In Fig.~\ref{fig:hysteresis}\textbf{B}, we show data for different mean detunings. The size of the hysteresis  region depends monotonously on the mean cavity detuning; the stronger the atoms are effectively coupled to the cavities, the larger is the hysteresis.

This hysteretic behavior reflects the discontinuous transition between the two self-organized phases. In the bistable regime, the atoms remain stably trapped in the configuration emitting in cavity 2, despite the fact that this configuration is now excited, until the coupling to the other cavity becomes sufficiently strong. Then, at a critical value of the detuning, the cavity mode populations abruptly change, signalling a quick reorganization into the pattern that scatters superradiantly into the other cavity mode 1 and that minimizes the energy. In the metastable pattern, in particular, the atomic configuration suppresses scattering into this mode, thus inhibiting the transition into the superradiant pattern. Due to these characteristics, we identify the metastable pattern as a subradiant state. 

The decay of the subradiant pattern into the superradiant phase proceeds via exponential amplification of the intracavity photon number in cavity 1. This instability reflects a non-equilibrium, driven-dissipative phase transition from subradiance to superradiance, with features reminiscent of dynamical phase transitions in systems with global-range interactions~\cite{Campa:2008}. Such subradiance-superradiance transitions have been theoretically proposed in various contexts~\cite{Gegg2018Superradiant,Shankar2021Superradiant,Hotter2023Cavity}, and our observations provide a direct manifestation of this mechanism under experimentally controlled conditions. In our case it is implemented by dispersively coupling the cavity with the external degrees of freedom of the quantum gas by means of the pump.

The numerical simulations of this dynamics use the GPE of Equation \eqref{Eq:GPE} and predict a hysteresis region that qualitatively reproduces the experimental measurement. As visible in Fig.~\ref{fig:hysteresis}\textbf{B}, the size of the theoretical hysteresis is however larger up to about a factor 5. We attribute this quantitative discrepancy to the model's inability to capture several experimental features. These include noise in the detunings $\Delta_\nu$, as well as a finite spatial overlap between the BEC and the two cavity mode functions, whose mode waists are located at slightly different spatial points, resulting in an inhomogeneous model and likely atomic currents during the ramp. The rates associated with contact interactions  and trap oscillations, in particular, are larger than the rate of the slow ramp, and are thus expected to influence the relaxation dynamics. This hypothesis is verified by comparison with numerical simulations that discard the contact interactions and the trap in the dynamics, hence setting $\omega_\nu=0$ and $V_0=0$ in Equation \eqref{Eq:GPE}. In this limit the GPE is reduced to a form known as Generalized GPE (GGPE) \cite{Plestid:2019}, where the sole interactions are of global range \cite{SM}. 
Full quantitative predictions from the GGPE are limited by its high sensitivity to initial conditions, such as density fluctuations seeded during the drive ramp-up. This sensitivity, combined with the difficulty of calibrating parameters across varying mean cavity detunings, hinders direct comparison with experimental data. Nonetheless, the GGPE consistently predicts smaller hysteresis regions than observed.
Since the GPE predicts substantially larger hysteresis regions than the GGPE, we identify trapping potential and interactions as stabilizing mechanisms that suppress the onset of large density fluctuations which would otherwise trigger radiative instabilities.

\begin{figure}[]
\begin{center}
\includegraphics[width=1\columnwidth]{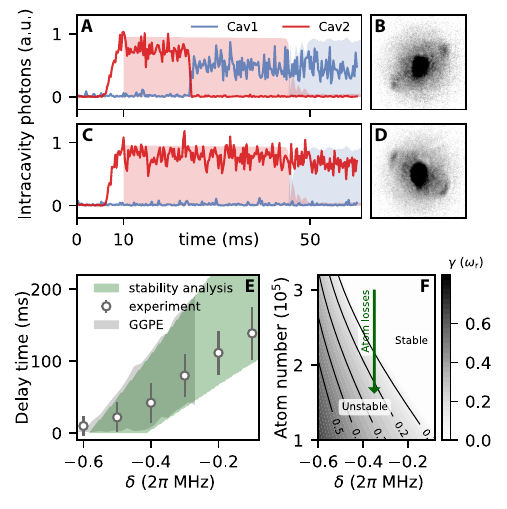}
\caption{
\textbf{Lifetime of subradiant states.} After preparing the system in a superradiant state of cavity 2 at $V_p=4.2\,\omega_r$, the relative detuning is quenched from $(\Delta_1,\Delta_2)=(\delta,0)$ to $(0,\delta)$. The system becomes subradiant with respect to cavity 1 until it eventually quickly reorganizes into the equilibrium, superradiant stripes with respect to cavity 1. 
\textbf{A}, \textbf{C} Two exemplary real-time traces of the cavity occupations for identical quench strength $\delta= 2\pi \times -0.4$~MHz: For the initial 10 milliseconds, cavity 2 is prepared in a superradiant state. After the quench, the system decays in the superradiant state of cavity 1 (panel \textbf{A}) or remains in the subradiant state for longer times (panel \textbf{C}). In the background (filled areas) we show a dynamical GGPE simulation for these parameters, initialized as superradiant state in cavity 2 at 10~ms. 
\textbf{B}, \textbf{D} Absorption images of the atomic momentum distributions at the end of the traces displayed in panels \textbf{A}, \textbf{C}, respectively.
\textbf{E} Delay time for which the system remains in the subradiant state, extracted from traces such as \textbf{A}, \textbf{C} as function of quench strength $\delta$. The grey area represents the GGPE simulation results, while the green area displays the results of our stability analysis. Both theory results span from the maximum atom number (minimum atom loss) to the minimum atom number (maximum atom loss). 
\textbf{F} Escape rates out of the subradiant state derived from the stability analysis as function of relative detuning and atom number. The estimated barrier height, Equation  \eqref{eq:Vbar}, remains constant along the solid lines. Atom loss eventually forces the system to leave the subradiant state.}
\label{fig:quenches}
\end{center}
\end{figure}

\paragraph*{Lifetime of Subradiant States.}
Subradiant states can also be prepared by suddenly tuning the cavity detunings across the phase diagram. In this way, a stable superradiant pattern becomes an excited, subradiant state. Experimentally, we initialize the system in the superradiant state with respect to cavity 2 by ramping up the drive field within 10~ms to 4.2~$E_r$ at fixed detunings $(\Delta_1,\Delta_2) = (\delta,0)$ and then abruptly swap detunings between the cavities to $(\Delta_1,\Delta_2) = (0,\delta)$. Quenching the phase is analogous to  extinguishing the driving field in subradiance experiments of laser-driven atoms \cite{Guerin2016Subradiance,Cipris2021Subradiance}. As shown in Fig.~\ref{fig:quenches}, we observe that the system keeps on emitting into cavity 2. The atoms still form the stripes that suppress emission into cavity 1, thus forming a subradiant phase.
Depending on the quench parameter $\delta$,  the BEC eventually reorders after a finite time into the superradiant state, emitting light exclusively into cavity 1.
We extract the lifetime of the subradiant state by fitting the time at which the cavity field occupations swap, see Fig.~\ref{fig:quenches}\textbf{A} and \textbf{C}. Fig.~\ref{fig:quenches}\textbf{E} shows the extracted lifetimes of the subradiant states as a function of the quench strength $\delta$. 

We identify four prominent features: (i) the lifetimes exceed any dynamic timescale of the system's constituents: The observed lifetimes are only compatible with the atom loss rate which is on the order of 3~Hz, see~\cite{SM}. (ii) The instability is characterized by a switch from the metastable subradiant to the stationary superradiant pattern, thereby preserving superfluidity, as can be seen from the absorption images in Fig.~\ref{fig:quenches}\textbf{B} and \textbf{D}. (iii) The transition itself is faster than the characteristic time scale set by trapping potential and the contact interaction. Moreover, it does not have features of nucleation. (iv) During the transition we observe emission in both cavity modes, even though the corresponding pattern of coexistence is energetically costly and thus unstable at the given pump strengths (see Fig.~\ref{fig:coexistence}). 

These features are characteristic for the out-of-equilibrium dynamics of strong long-range interacting systems. The short-time behavior of the transition exhibits the features of a violent relaxation, characteristic of Vlasov instabilities in plasmas and gravitational systems \cite{Balescu:1960,Henon:1982,Bertrand:2019,Elskens:2020}.
These instabilities have a characteristic exponential amplification of fluctuations \cite{Lynden-Bell:1967,Henon:1982,Campa:2009} and can be triggered by a change of the energy \cite{Jain:2007,Campa:2008} or by lowering the barrier of the confining potential below the threshold \cite{Chavanis:metastable,Saadat:2023}. In our case the barrier is the depth of the mechanical potential confining the stripes, see Eq.\ \eqref{eq:Vbar}: The nonlinear dependence on $N$ shows that atom losses tend to ramp it down. When the barrier is lowered to $\bar V\sim \omega_r$, the atoms flow out from the subradiant stripes at the intersection points with the superradiant stripes, filling these quickly. During this process the atoms form a pattern that supports simultaneous emission into both cavity modes. The instability is seeded by fluctuations of the atoms' kinetic energy in the repulsive potential. However, we are experimentally unable to identify wether the fluctuations seeding the instability are quantum or thermal. The long-lived metastable state is superfluid and in this sense it is an experimental example of a {\it quantum} metastable state due to long-range interactions. The reorganization process preserves superfluidity since the long-range interaction term of Eq.\,\eqref{Eq:GPE} does not affect off-diagonal long-range order.

The instability is a violent relaxation and the process of filling the superradiant stripes via a transient excited configuration minimizes the long-range interactions. This dynamics is captured by the GGPE, which in turn describes a quantum version of the so-called generalized mean-field model \cite{Teles:2012}.
The analysis of the linear fluctuations about the equilibrium state of the GGPE shows that the subradiant phase is a fixed point and permits to describe its decay process. The linear fluctuations $\delta\psi$ about the subradiant state $\psi_0$ are governed by the equation \cite{SM}
\begin{equation}
    \hbar\omega\delta\psi=\left(\frac{\hbar^2\nabla^2}{2m}-H_{\rm mf}[\psi_0]\right)\delta\psi-\delta H_{\rm mf}[\psi_0]\psi_0\,,\label{eq:FT_deltapsi}
\end{equation}
where $H_{\rm mf}[\psi_0]$ and $\delta H_{\rm mf}[\psi_0]$ are, respectively, the mean-field Hamiltonian and its variation around the subradiant state, both depending on the atom number $N$. The instability is signalled by ${\rm Im}\omega<0$, allowing us to identify a threshold value $N_c$ above which the subradiant state is a stable solution of the GGPE.
As illustrated in Fig.~\ref{fig:quenches}\textbf{F}, it eventually becomes unstable because atom losses slowly decrease the atom number until $N < N_c$. The system then quickly escapes out of the subradiant state with exponential growth at rate $\gamma=-{\rm Im}\omega$, typical of violent relaxation. The results of this analysis are reported in Fig.\ \ref{fig:quenches}\textbf{E} and are in agreement with the experimental data, showing that this simplified model captures the essential features both qualitatively as well as quantitatively. 

While Equation \eqref{eq:FT_deltapsi} describes the short-time dynamics of the subradiant-superradiant transition, the full transition is captured by the GGPE. Fig.~\ref{fig:quenches}~\textbf{A},\textbf{C} shows the agreement between the photon traces of experiment and theory within the uncertainty in the atom loss rate. We have compared these predictions with the simulations using the full GPE, which again includes contact interactions and trap potentials, and found that the lifetime of the subradiant state results to be larger, yet within the same order of magnitude. This confirms that the relaxation dynamics (different than the system's response to slow ramps when measuring the hysteresis region) is primarily governed by the interplay of (quantum) fluctuations, initiating the transition, and long-range interactions, driving the system into the stable, superradiant state. In both descriptions, the final state aligns with experimental observations, demonstrating that long-range interactions dominate the overall dynamics across the subradiant-superradiant transition. 

\section*{Discussion}
We have investigated the relaxation dynamics of a quantum Bose gas with competing global-range interactions mediated by photons. These interactions impose two competing spatial patterns, leading to a phase diagram  which we have explored by tuning the cavity frequencies. The transition between the two configurations is of first order. A slow ramp across that transition reveals a typical hysteretical behavior: within the hysteresis region the gas remains trapped in the pattern that suppresses emission in the other cavity, forming a subradiant grating. At the instability, it quickly reorders into the superradiant grating undergoing a non-equilibrium phase transition from a subradiant to a superradiant pattern. Energy considerations show that the reordering mechanism is dominated by the global-range interactions. Our theoretical study indicates that the breadth of the hysteresis region can be reproduced quantitatively only by taking into account fine-tuned contact interactions and spatial inhomogeneities.

The relaxation following sudden quenches across the phase diagram, instead, is solely governed by the interplay of global forces and kinetic energy. We identified parameter regimes where the subradiant patterns are excited metastable states. Their lifetime is limited only by atom losses which slowly reduce the strength of the interactions, and thus the depth of the barrier trapping the subradiant state. 
This dynamics is paradigmatic of collective phenomena of systems with strong long-range interactions, such as clustering, phase transitions, and fragmentation, characteristic of stars in a cluster or particles in a plasma, whose onset or stability strongly depends on the number of particles involved~\cite{Campa:2009,Chavanis2005On,Chavanis2006Phase}.
Instabilities in strong-long range interacting systems due to atom losses have also been discussed in the literature of ideal models \cite{Gupta:2010}, for which our many-body cavity QED setup provides a testbed.

The observed instability is seeded by density fluctuations due to the kinetic energy and exhibits characteristics of a violent relaxation, characteristic of the dynamical instability of mean-field, nonlinear equations, such as the Vlasov equation. Differing from experiments with optical systems \cite{Faccio:2024}, the instability  emerges from the many-body dynamics of ultracold Bosons that interact via multiple photon scattering. Instability and reorganization in the superradiant pattern are captured by the interplay of global interactions and kinetic energy. Despite the change of energy at the transition from subradiance to superradiance, the superfluid properties of the quantum gas are preserved. The model can be mapped to a quantum realisation of the paradigmatic generalized mean-field model, which predicts a plethora of non-equilibrium phase transition from the the interplay of competing long-range forces \cite{Teles:2012,Pikovsky:2014,Silva:2020,Keller:2018}. The cavity QED dynamics thus allows to shed light on universal equilibrium and out-of-equilibrium critical phenomena due to frustration between globally-interacting potentials. Our findings can also be connected to the non-equilibrium dynamics of ultracold Fermions with photon-mediated interactions in vicinity of a second-order phase transition. There, relaxation to a stationary state was observed within less than a millisecond, but with typical features of a dynamical instability~\cite{Wu2023Signatures,zwettler2024nonequilibrium,marijanovic2024dynamical}. Analogous dynamics have been predicted for the relaxation of thermal atoms in cavities \cite{Schuetz:2014}. Such characteristics are thus distinctive of instabilities in long-range interacting systems across various length scales and in this sense universal. 

Our results shed light on the individual mechanisms at work and on their interplay. By establishing a direct connection between the microscopic dynamics and the dynamical instability of a nonlinear equation for few macroscopic variables, which reproduce the experimental observations, we provide an example of universal behavior in the out-of-equilibrium dynamics. Our work demonstrates the potential of many-body cavity quantum electrodynamics to unveil the processes leading to relaxation in long-range interacting systems. Moreover, by validating the  effective nonlinear model derived for the out-of-equilibrium dynamics, this study lays the foundation for concepts and strategies to control metastable states in quantum many-body systems.

\acknowledgements
We acknowledge stimulating discussions with Immanuel Bloch, Tilman Esslinger, B.-W.\ Himmelreich, Stefan Ostermann, and Susanne Yelin, and thank Philip Zupancic for early contributions to the project. Furthermore we thank Christiane Koch, Jean Dalibard and Ana Maria Rey for valuable input on our manuscript.

TD acknowledges support by Swiss National Science Foundation (SNSF), SNSF project numbers 217124, 221538, and 186312, and support from the Swiss State Secretariat for Education, Research and Innovation 
(SERI) (project number MB22.00090).
GM acknowledges support by the Deutsche Forschungsgemeinschaft (DFG, German Research Foundation) -- DFG project ID 429529648 -- TRR306 QuCoLiMa ("Quantum Cooperativity of Light and Matter"), and the DFG project "Quantum many-body dynamics of matter and light in cavity QED" - DFG project ID 525057097, and the QuantERA project "QNet: Quantum transport, metastability, and neuromorphic applications in Quantum Networks" - DFG project ID 532771420. GM further acknowledges support by the National Science Foundation under Grants No. NSF PHY-1748958 and PHY-2309135.
CM acknowledges support from the Conselho Nacional de Desenvolvimento Científico e Tecnológico, CNPq project No 201765/2020-9.




%

\newpage
\widetext
\begin{appendix}
\begin{center}
\textbf{\large Supplemental Materials}\label{SM}
\end{center}

\section*{Theoretical Model: Gross-Pitaevskii equation with global-range interactions}
    In this Appendix, we provide details about the theoretical model of the system dynamics and the numerical simulations we performed, leading to the theoretical data presented in the main text. We first introduce the full quantum mechanical model of atoms and cavities. By adiabatically eliminating the cavities' degrees of freedom and employing a mean-field ansatz for the state of the atomic sector, we derive in the subsequent sections the mean-field Hamiltonian of Eq.\ (2) in the main text. At the end, we give additional information about the numerical simulation of the GPE, Eq.\ (1) of the main text, for the various protocols of the experiment.

    \subsection*{Quantum master equation for the composite system}
    In what follows, we denote by $\hat{a}_\nu$ and $\hat{a}_\nu^\dag$ the annihilation and creation operators of cavity mode $\nu$ at frequency $\omega_\nu^c$ and wave vector $\bm{k}_\nu$, obeying the bosonic commutation relations $[\hat{a}_\nu,\hat{a}_\mu^\dagger] = \delta_{\nu,\mu}$ and $[\hat{a}_\nu,\hat{a}_\mu] = 0$. For convenience, we treat the atoms in first quantization and denote by $\hat{\bm{x}}_j$ and $\hat{\bm{p}}_j$ the canonically-conjugate positions and momenta in the $x-y$ plane. A given atoms' dipolar transition at frequency $\omega_a$ couples with both cavity modes and with a transverse classical field at frequency $\omega_p$ and wave vector $\bm{k}_p$. Below we assume that the detuning  $\Delta_a=\omega_p-\omega_a$ of the laser from the atomic transition is the largest frequency, allowing us to eliminate the atoms' internal degrees of freedom from the equation of cavity and atomic external degrees of freedom and to discard spontaneous emission~\cite{Schutz2013}
    . This requires that the spontaneous decay rate $\gamma\ll|\Delta_a|$. It further requires that the laser Rabi frequency $\Omega\ll |\Delta_a|$ and that the vacuum Rabi frequencies satisfy $g_\nu\sqrt{N}\ll |\Delta_a|$. As in the experiment, we will take $\Delta_a>0$. 
    After eliminating the electronic degrees of freedom, we obtain the quantum master equation governing the dynamics of the density operator $\hat\rho$, which describes the state of the two cavity modes ($\nu = 1,2$) and of the motional degrees of freedom of the particles in the reference frame rotating at the laser frequency $\omega_p$~\cite{Schutz2013,Zupancic2019P}
    : 
    \begin{equation}
        \frac{\partial}{\partial t}\hat{\rho} = -\frac{i}{\hbar}\left[\hat{H},\hat{\rho}\right] + \sum_{\nu=1}^2\kappa_\nu\mathcal{L}[\hat{a}_\nu]\hat{\rho}\label{app:full_master_eq}\,.
    \end{equation}
    The cavity modes' losses at rate $\kappa_\nu$ are accounted for by the last term, where $\mathcal{L}[\hat{a}_\nu]\hat{\rho}=2\hat{a}_\nu\hat{\rho}\hat{a}_\nu^\dagger -\hat{a}_\nu^\dagger \hat{a}_\nu\hat{\rho} -\hat{\rho}\hat{a}_\nu^\dagger \hat{a}_\nu$. The (effective) Hamiltonian $\hat{H}=\hat{H}_\mathrm{a} + \hat{H}_\mathrm{f} + \hat{V}_\mathrm{int}$ is the sum of the coherent free dynamics of the $N$ particles ($\hat{H}_\mathrm{a}$), of the  cavity modes ($\hat{H}_\mathrm{f}$), and of their optomechanical interactions $\hat{V}_\mathrm{int}$:
    \begin{eqnarray}
        \hat{H}_\mathrm{a} & =& \sum_{j=1}^N\left(\frac{\hat{\bm{p}}^2_j}{2m} + \hbar V_p\cos^2(\bm{k}_p\cdot \hat{\bm{x}}_j)\right)\,,\\
        \hat{H}_\mathrm{f} &=& -\sum_{\nu = 1}^2 \hbar \Delta_\nu \hat{a}_\nu^\dagger \hat{a}_\nu\,,\\
        \hat{V}_\mathrm{int} &=& \sum_{\nu = 1}^2\hbar U_\nu \hat{\Theta}_{\nu\nu}\hat{a}_\nu^\dagger\hat{a}_\nu + \sum_{\nu = 1}^2\hbar \sqrt{V_p U_\nu}\hat{\Theta}_{p\nu}(\hat{a}_\nu + \hat{a}_\nu^\dagger) + \hbar \sqrt{U_1U_2}\hat{\Theta}_{12}(\hat{a}_1^\dagger\hat{a}_2 + \hat{a}_2^\dagger\hat{a}_1)\,.
    \end{eqnarray}
    where $V_p = \Omega^2/\Delta_a$ is the pump lattice depth and $U_{\nu} = g_{\nu}^2/\Delta_a$ is the single-particle dispersive shift due to the coupling with the cavity mode~\cite{Mivehvar2021Cavity}
    . We used the atomic operators
    \begin{equation}
        \hat{\Theta}_{\mu\nu} = \sum_{j=1}^N\cos(\bm{k}_\mu \cdot \hat{\bm{x}}_j)\cos(\bm{k}_\nu \cdot \hat{\bm{x}}_j)\,,\quad \mu,\nu = p,1,2\,,
    \end{equation}
    where $\bm{k}_1$ and $\bm{k}_2$ are the wavevectors of the cavities. The slow experimental ramps of laser intensity and the cavity frequencies are described by treating $V_p$ and $\Delta_\nu$ as time-dependent coefficients in \eqref{app:full_master_eq}.
    
    \subsection*{Adiabatic elimination of the cavities degrees of freedom}
    In order to simplify the theoretical description of the system, we employ the method developed in Refs.~\cite{Jaeger2022,Schmit:2024}
    to adiabatically eliminate the cavities' degrees of freedom from the master equation (\ref{app:full_master_eq}). This leads to an effective master equation, which we here report for the specific choice $\Delta_a > 0$:
    \begin{equation}
        \frac{\partial}{\partial t}\hat{\rho}_\mathrm{sys} = -\frac{i}{\hbar}\left[\hat{H}_\mathrm{a} + \frac{\hbar}{2}\sum_{\nu=1}^2\sqrt{V_p U_\nu}\left(\hat{\alpha}_\nu^\dagger \hat{\Theta}_{p\nu} + \hat{\Theta}_{p\nu}^\dagger\hat{\alpha}_\nu\right),\hat{\rho}_\mathrm{sys}\right] + \sum_{\nu=1}^2\kappa_\nu \mathcal{L}[\hat{\alpha}_\nu]\hat{\rho}_\mathrm{sys}\label{app:effective_master_equation}\,,
    \end{equation}
    describing the dynamics of the particles' external degrees of freedom, with $\hat{\rho}_\mathrm{sys}$ the reduced atomic density operator. The atomic operators $\hat{\alpha}_{1,2}$ represent effective quantum fields~\cite{Larson:2008}
    and they obey the coupled equations of motion
    \begin{align}
        \frac{\partial}{\partial t}\hat{\alpha}_1 &=-\frac{i}{\hbar}[\hat H_\mathrm{a},\hat\alpha_1] -i\hat{Z}_1\hat{\alpha}_1
	-i\sqrt{V_p U_1} \hat{\Theta}_{p1} - i \sqrt{U_1U_2} \hat{\Theta}_{12} \hat{\alpha}_2\,,\\
 \frac{\partial}{\partial t}\hat{\alpha}_2 &=-\frac{i}{\hbar}[\hat H_\mathrm{a},\hat\alpha_2] -i\hat{Z}_2\hat{\alpha}_2
	-i\sqrt{V_p U_2} \hat{\Theta}_{p2} - i \sqrt{U_1U_2} \hat{\Theta}_{12} \hat{\alpha}_1\,.
    \end{align}
    When the time scales of the cavity mode dynamics are much shorter than the ones characterizing the motion, we can replace the operators $\hat{\alpha}_{1,2}$ in \eqref{app:effective_master_equation} by their adiabatic solutions, see Refs~\cite{Larson:2008,Schutz2013}.
    The latter are found from the previous equations by setting $\partial_t\hat{\alpha}_{1,2}=0$ and neglecting the kinetic energy during the cavity relaxation to the local steady state, and take the form:
    \begin{eqnarray}
        \hat{\alpha}_1 = \sqrt{V_pU_1}\frac{\hat{\Theta}_{p1}\hat{Z}_2 - U_2 \hat{\Theta}_{12}\hat{\Theta}_{p2}}{U_1U_2\hat{\Theta}_{12}^2 - \hat{Z}_1\hat{Z}_2}\quad \text{and} \quad \hat{\alpha}_2 = \sqrt{V_p U_2}\frac{\hat{\Theta}_{p2}\hat{Z}_1 - U_1 \hat{\Theta}_{12}\hat{\Theta}_{p1}}{U_1U_2\hat{\Theta}_{12}^2 - \hat{Z}_1\hat{Z}_2}\label{eq:alpha_1_and_alpha_2_op}\,,
    \end{eqnarray}
where $\hat{Z}_\nu = -\Delta_\nu + U_\nu \hat{\Theta}_{\nu\nu} - i\kappa_\nu$, with $\nu = 1,2$.

\subsection*{Mean-field Hamiltonian}

    The mean-field master equation is derived using the mean-field ansatz
    \begin{equation}
        \hat{\rho}_\mathrm{sys} = \otimes_{j=1}^N \hat{\rho}_a\,.
    \end{equation}
    Integrating out the other $N-1$ variables we obtain the (non-linear) mean-field master equation
    \begin{equation}
        \frac{\partial}{\partial t}\hat{\rho}_a = -\frac{i}{\hbar}\left[\frac{\hat{\bm{p}}^2}{2m} +\hat{H}_\mathrm{mf}[\hat{\rho}_a],\hat{\rho}_a\right]\label{eq:mf_master_equation}\,,
    \end{equation}
    describing the dynamics of the single-particle density operator $\hat{\rho}_a$. The mean-field Hamiltonian takes the form 
    \begin{equation}
        \begin{split}
        \hat{H}_\mathrm{mf}[\hat{\rho}_a] &=  \hbar V_p\hat{\theta}_p^2 + \sum_{\nu = 1}^2\hbar U_\nu |\alpha_\nu[\hat{\rho}_a]|^2\hat\theta_\nu^2  + \sum_{\nu = 1}^2 2\hbar \sqrt{V_p U_\nu}\mathrm{Re}(\alpha_\nu[\hat{\rho}_a])\hat{\theta}_p\hat\theta_\nu\\ 
        &+ 2 \hbar \sqrt{U_1 U_2}\mathrm{Re}(\alpha_1^*[\hat{\rho}_a]\alpha_2[\hat{\rho}_a])\hat{\theta}_1\hat\theta_2\label{eq:mf_Hamiltonian}\,.
        \end{split}
    \end{equation}
    
    Here, $\hat{\theta}_\mu = \cos(\bm{k}_\mu \cdot \hat{\bm{x}})$, with $\mu = p,1,2$, and the amplitudes of the cavity fields $\alpha_{1,2}=\langle\hat\alpha_{1,2}\rangle\equiv {\rm Tr}\{\hat\alpha_{1,2}\hat\rho_{\rm sys}\}$, with the operators $\hat{\Theta}_{\mu\nu}$ essentially replaced by the expectation values $N\langle\hat{\theta}_{\mu}\hat{\theta}_{\nu}\rangle$.
Their expressions are found by using the steady-state equations \eqref{eq:alpha_1_and_alpha_2_op} and using the mean-field approximation $\langle \hat{\Theta}_{\mu\nu} \hat{\alpha}_{1,2} \rangle \approx N\langle \hat{\theta}_\mu \hat{\theta}_\nu\rangle \langle\hat{\alpha}_{1,2} \rangle$:
    \begin{equation}
    \begin{split}
        \alpha_1[\hat{\rho}_a] &= \sqrt{V_p U_1}N\frac{\langle\hat{\theta}_{p}\hat{\theta}_{1}\rangle Z_2 - U_2 N \langle\hat{\theta}_{p}\hat{\theta}_{2}\rangle\langle\hat{\theta}_{1}\hat{\theta}_{2}\rangle}{U_1 U_2 N^2\langle\hat{\theta}_{1}\hat{\theta}_{2}\rangle^2-Z_1Z_2}\quad \text{and} \quad \\
        \alpha_2[\hat{\rho}_a] &= \sqrt{V_p U_2 }N\frac{\langle\hat{\theta}_{p}\hat{\theta}_{2}\rangle Z_1 - U_1 N \langle\hat{\theta}_{p}\hat{\theta}_{1}\rangle\langle\hat{\theta}_{1}\hat{\theta}_{2}\rangle}{U_1 U_2 N^2\langle\hat{\theta}_{1}\hat{\theta}_{2}\rangle^2-Z_1Z_2}\label{eq:steady-state_solutions_alpha}\,,
    \end{split}
    \end{equation}	
    where $Z_{\nu=1,2} = U_\nu N \langle \hat{\theta}_\nu^2\rangle - \Delta_{\nu}-i\kappa_\nu$. Hamiltonian $\hat{H}_\mathrm{mf}$ depends nonlinearly on the atomic state through the functionals $\alpha_{1,2}$. The latter can be rewritten as
\begin{eqnarray}
\alpha_1[\hat{\rho}_a] &=& -\frac{N\sqrt{V_pU_1}}{Z_1}\frac{\langle\hat{\theta}_{p}\hat{\theta}_{1}\rangle + C_2' \langle\hat{\theta}_{p}\hat{\theta}_{2}\rangle\langle\hat{\theta}_{1}\hat{\theta}_{2}\rangle}{1-C_1'C_2'\langle\hat{\theta}_{1}\hat{\theta}_{2}\rangle^2}\label{eq:steady-state_solutions_alpha:1}\,,
\end{eqnarray}
where $C_\nu'=-NU_\nu/Z_\nu$. Note that $C_\nu'\to C_\nu$ for negligible cavity decay rate.

For vanishing cavity loss rates (namely, when $\alpha_\nu$ is real), the mean-field Hamiltonian (\ref{eq:mf_Hamiltonian}) can be cast in the form
    \begin{eqnarray}
        \hat{H}_\mathrm{mf}^{\rm (blue)} &=& \hbar \left(\sqrt{V_p}\hat{\theta}_p + \sum_{\nu = 1}^2\sqrt{U_\nu} \alpha_\nu\hat\theta_\nu\right)^2  \label{eq:mf_Hamiltonian:blue}\,,
    \end{eqnarray}
    thus showing that the mean-field energy is bounded from below, $\langle\hat{H}_\mathrm{mf}^{\rm (blue)}\rangle\ge 0$. This expression is strictly valid provided that $V_p$, $U_{\nu=1,2}$ are positive, corresponding to an atom-pump detuning on the blue side, $\Delta_\mathrm{a}>0$. For red detunings, instead, these quantities are negative and the mean-field Hamiltonian can be rewritten as 
\begin{eqnarray}
        \hat{H}_\mathrm{mf}^{\rm (red)} &=& -\hbar \left(\sqrt{|V_p|}\hat{\theta}_p + \sum_{\nu = 1}^2\sqrt{|U_\nu|} \alpha_\nu\hat\theta_\nu\right)^2 \label{eq:mf_Hamiltonian:red}\,.
\end{eqnarray}

\subsection*{Blue self-organization, metastability, and threshold for phase coexistence}
Opposite to the red-detuned case, where the mean-field energy is always negative, for blue detuning the mean-field energy is positive. This is an essential feature, that leads to strikingly different equilibrium properties. We summarize some here, and present the underlying theory elsewhere~\cite{Schmit:2024}.
Self-organization in the blue takes place within the detuning range $\Delta_\nu\in [-NU_\nu/2,NU_\nu/2]$, where the specific size of the interval varies as a function of $V_p$~\cite{Zupancic2019P}.
In addition, when the two cavity mode wave vectors are neither parallel nor orthogonal, we observe that simultaneous self-organization in both modes (phase coexistence) occurs at a larger pump threshold than for self-organization in one of the modes, even for $\Delta_1\approx\Delta_2$. This is also in contrast with self-organization for red detunings, where the different orders typically coexist and can even mutually enhance each other~\cite{Morales2017Coupling}. 
Below we provide a simple argument for the existence of two thresholds for the pump, a lower one, $V_p^{\mathrm{th}_\nu}$, separating the disordered phase from self-organization in one mode and a higher one,$V_p^\mathrm{th_{12}}$, separating single-mode ordering from phase coexistence.

Minimization of the mean-field energy, $\langle \hat{H}_\mathrm{mf}^{\rm (blue)}\rangle=0$, is trivially achieved when the atoms are at the nodes of the pump lattice and the cavity modes are empty. 
This becomes mechanically unstable for increasing $V_p$, and the new stationary state is characterized by patterns such that one or both cavity fields do not vanish but destructively interfere with the pump field at the atomic positions. For blue-detuned self-organization, we can assume a maximal overlap of $1/2$ between the atomic density and any of the potentials, $\langle \hat{\theta}_{\nu=1,2}^2\rangle=\langle \hat{\theta}_p^2\rangle=\langle \hat{\theta}_{\nu}\hat{\theta}_p\rangle=1/2$, since density stripes can form in the direction determined by either $\bm{k}_p+\bm{k}_\nu$ or $\bm{k}_p-\bm{k}_\nu$. Specifically, also phase coexistence requires $\langle \hat{\theta}_1\hat{\theta}_2\rangle=1/2$.
Using these relations, the mean-field energy vanishes when the amplitudes of the cavity field modes satisfy the relation:
\begin{equation}
\label{eq:blue:interference}
    \sqrt{V_p} = -\left(\sqrt{U_1}\alpha_1 + \sqrt{U_2}\alpha_2\right)\,,
\end{equation}
which can be exactly fulfilled for $NU_\nu\gg |\Delta_\nu|$~\cite{Zippilli_2004}.

We first determine the threshold $V_p^{\mathrm{th}_\nu}$ on the pump for self-organization in one cavity mode and assume that the atoms form stripes at the wave vector $\bm{k}_p+\bm{k}_1$, such that $\alpha_1=-\sqrt{V_p/U_1}$ and $\alpha_2=0$. The stripes are confined by the potential $\bar{V}\cos((\bm {k}_p+\bm{k}_1)\cdot\bm{x})$, with depth 
$\bar{V}=\sqrt{V_p U_1}\alpha_1$, and are mechanically stable when the potential height is larger than the minimal amount of mechanical energy transferred to an atom by photon scattering, i.e. when $|\bar{V}|>{\mathcal C}\omega_r$, where $\mathcal C\gtrsim1$ is a constant. Therefore, the stripes' mechanical stability requires $|\alpha_1|\ge \mathcal C\omega_r/\sqrt{V_pU_1}$. The explicit dependence of the amplitude $\alpha_1$ on $V_p$ is given in \eqref{eq:steady-state_solutions_alpha:1}. For $\alpha_2=0$ then
\begin{equation}
\label{eq:Vbar}
    \bar V =  {V_p}\frac{U_1 N}{U_1 N/2 - \Delta_1}\,.
\end{equation}
and the stripe is (meta)stable when the pump exceeds the threshold value $V_p^{\mathrm{th}_\nu} = V_p^{\mathrm{th}_1}$, with $V_p^{\mathrm{th}_1}\equiv \mathcal{C}\omega_r (U_1N/2-\Delta_1)/(U_1N)$ a function of the detuning $\Delta_1$.
Note that $V_p^{\mathrm{th}_1}\le \mathcal C\omega_r$.
The depth of the potential confining the stripes, $\bar V$, is also the barrier that separates the subradiant from the superradiant pattern. The subradiant pattern, in particular, becomes unstable when the number of atoms reaches the value such that $\bar V\le \mathcal C\omega_r$. 

Relation (\ref{eq:blue:interference}) leads to a constraint for the maximum value that each field amplitude can reach, $|\alpha_\nu|\le \sqrt{V_p/U_\nu}$, see Ref~\cite{Zippilli_2004}, 
and constrains at the same time the maximum value of the second cavity field mode, given a certain field in the first cavity. This allows us to estimate a lower bound $V_p^{\mathrm{th}_{12}}$ for the pump threshold for phase coexistence. Starting from \eqref{eq:blue:interference}, self-organization in both cavity modes requires that both (negative) cavity field amplitudes satisfy $|\alpha_\nu|>\mathcal{C}\omega_r/\sqrt{U_\nu V_p}$, leading to the inequality $\sqrt{V_p} \ge 2\mathcal C\omega_r/\sqrt{V_p}$. The value of $V_p= V_p^{\mathrm{th}_{12}}$ at which the equality holds is the threshold for phase coexistence, and reads:
\begin{equation}
\label{eq:blue:interference:th}
    V_p^{\mathrm{th}_{12}}= 2\mathcal C\omega_r\,.
\end{equation}
This is at least twice as large as the threshold for stripe formation $V_p^{\mathrm{th}_\nu}$ and qualitatively agrees with the numerical simulation in Fig.~2\textbf{D}. 

\subsection*{Gross-Pitaevskii and Generalized Gross-Pitaevskii equation}

    The Gross-Pitaevskii equation (GPE), Eq.~(1) in the main text, is derived from \eqref{eq:mf_Hamiltonian} after writing the mean-field Hamiltonian in second quantization, adding the trapping potential and the van-der-Waals scattering term, and making the ansatz of symmetry breaking. The generalized Gross-Pitaevskii equation (GGPE) follows directly from the GPE by neglecting the trapping potential, $\omega_{\nu =x,y} = 0$, and the short-range interaction, $V_0 = 0$:
    \begin{equation}
        i\hbar \frac{\partial}{\partial t}\psi(\bm{x},t) = \left(-\frac{\hbar^2 \nabla^2}{2m} + H_\mathrm{mf}[\psi]\right)\psi(\bm{x},t)\label{eq:GGPE}\,.
    \end{equation}
    The mean-field Hamiltonian $H_\mathrm{mf}$ is identical to \eqref{eq:mf_Hamiltonian}, where the expectation values in the cavity amplitudes, \eqref{eq:steady-state_solutions_alpha}, are now taken over the wavefunction $\psi$, $\langle \hat{\theta}_\mu \hat{\theta}_\nu \rangle = \int \mathrm{d}\bm{x}~\psi^*(\bm{x},t)\cos(\bm{k}_\mu \cdot \bm{x})\cos(\bm{k}_\nu \cdot \bm{x})\psi(\bm{x},t)$.
    
    \subsection*{Numerical Simulations}
    In this section, we give further details on the numerical simulations we performed based on the GPE and the GGPE.
    
    \subsubsection*{Implementation}
    The GPE has been implemented numerically using the Julia package QuantumOptic.jl~\cite{kramer2018quantumoptics}. 
    The wavefunction $\psi(\bm{x},t)$ is here discretized in space over a two-dimensional grid $[-L_x/2,L_x/2] \times [-L_y/2,L_y/2]$ with $n_x \times n_y$ points. The lengths $L_{x,y}$ are fixed by the extension of the atomic cloud, which can be estimated by the spatial extension of a trapped two-dimensional BEC in its ground state. In the Thomas-Fermi limit, the ground-state density is close to the Thomas-Fermi solution $|\psi_\mathrm{TF}|^2$, found by solving the GPE (see Eq. (1) of the main text) for its steady state when neglecting the mean-field potential $H_\mathrm{mf}$ and the kinetic energy term:
    \begin{equation}
        |\psi_\mathrm{TF}(x,y)|^2 = \frac{2}{\pi R_x R_y}\begin{cases}
            1 - \left(\frac{x}{R_x}\right)^2 - \left(\frac{y}{R_y}\right)^2,& (x/R_x)^2 + (y/R_y)^2 \leq 1\\
            0,&\text{otherwise}\,.
        \end{cases}
    \end{equation}
    The size of the cloud is characterized by the Thomas-Fermi radii
    \begin{equation}
        R_{\nu} = \frac{1}{\omega_\nu}\left(\frac{4}{m\pi}V_0 N \omega_x \omega_y\right)^\frac{1}{4}\,,\quad \nu = x,y\,.
    \end{equation}
    Typical grid sizes $L_{x,y}$ are chosen between $1.4-1.6$ times the Thomas-Fermi radii. While the extension of the cloud is mainly dictated by the short-range interaction and the harmonic trapping potential, the step sizes $L_x/n_x$ and $L_y/n_y$ are controlled by the light fields of the pump and the cavities, in particular, the periodic potentials they create. For the considered values of the cavity detunings $\Delta_{1,2}$, we can assume that all potentials are chracterized by the same wavenumber $k = \sqrt{2m\omega_r/\hbar} \approx 8 ~\mu \mathrm{m}^{-1}$, with $\omega_r$ the recoil frequency. In the simulations, we choose the stepsizes to fit $8-12$ grid points within one wavelength $\lambda = 2\pi/k$ to resolve these structures. For the experimental parameters, this requires typically a grid of $n_{x,y} = 200$ - $400$ points per spatial direction.

    It is worth noting that for solving the GGPE of \eqref{eq:GGPE}, we can exploit the periodicity of the mean-field Hamiltonian (\ref{eq:mf_Hamiltonian}) and reduce the spatial grid to a single unit cell, as confirmed by the numerics. The unit cell is determined by the wavevectors of the pump and cavity modes, given for the specific geometry of our experimental setup by $\bm{k}_p = k\bm{e}_y$, $\bm{k}_1 = k(\sqrt{3}/2\bm{e}_x -1/2\bm{e}_y)$, and $\bm{k}_2 = k(\sqrt{3}/2\bm{e}_x + 1/2\bm{e}_y)$. Here, $\bm{e}_x$ and $\bm{e}_y$ are the unit vectors of the two spatial directions and, without loss of generality, we align the $y$-axis with the pump axis. The dimensions of the unit cell are here $L_x = (4\pi/\sqrt{3})k^{-1}$ and $L_y = 4\pi k^{-1}$.

    \subsubsection*{Contact-interaction potential}
    In three dimensions, the contact interaction is given by $V_0^{(\mathrm{3D})} = 4\pi \hbar^2 a_\mathrm{s}/m$~\cite{Pitaevskii_Stringari_BEC}. 
    For Rubidium $87$, the mass is $m = 1.44316\times 10^{-25}$ kg
    and the scattering length is $a_\mathrm{s} = 98 a_0$, with $a_0$ the Bohr radius. In three dimensions, the Thomas-Fermi radii can be derived similarly to the preceding paragraph, leading to
    \begin{equation}
        R_\nu^{(\mathrm{3D})} = \frac{1}{\omega_\nu}\left(\frac{15}{4\pi}\frac{V_0^{(\mathrm{3D})} N \omega_x \omega_y \omega_z}{m}\right)^\frac{1}{5}\,.
    \end{equation}
    The two-dimensional contact interaction strength $V_0$ in our GPE simulations is chosen so that the extension of the cloud in the $x$-$y$-plane matches the one of a three-dimensional cloud. Thus, by requiring that $R_{x,y}^{(\mathrm{3D})} = R_{x,y}$, we find the following expression for the two-dimensional contact interaction strength
    \begin{equation}
        V_0 = \left(\frac{15^4 \pi}{2^{18}}\frac{m \omega_z^4}{\omega_x \omega_y} \frac{\left(V_0^{(\mathrm{3D})}\right)^4}{N}\right)^\frac{1}{5}\label{Eq:2D_V0_definition}\,.
    \end{equation}

    \subsubsection*{Normalization and atom losses}
    The condensate wavefunction $\psi$ is here defined such that the normalization $\mathcal{P} =\int\mathrm{d}\bm{x}~|\psi(\bm{x})|^2 = 1$ is independent of the number of particles $N$, as such, $N$ appears explicitly in the GPE via the short- and long-range interactions (see, for instance, \eqref{eq:steady-state_solutions_alpha}). When solving the GPE for the dynamics of the system, we account for the atom losses of the experiment. This is done by adding to the GPE a phenomenological loss term $-i\Gamma/2\psi$, leading to an exponential decay of the initially normalized wavefunction over time with rate $\Gamma$, such that $\mathcal{P}(t) = e^{-\Gamma t}$. As extracted from the characterization of the experiment (see Sec. "Atomic loss rate"), we estimate the loss rate $\Gamma$ to range between $2$ and $4$ Hz. In the presence of atom losses, $N$ represents in the dynamical equation the initial number of atoms in the BEC, while the number of atoms at time $t$ is given by $N\mathcal{P}(t)$. It is important to note that the contact interaction strength \eqref{Eq:2D_V0_definition} has to be replaced by $V_0 e^{\Gamma t/5}$ in the GPE to ensure the correct size of the cloud in the $x$-$y$-plane (see Sec. 
    "Contact-interaction potential") at any time $t$.

\section*{Linear Stability Analysis}
In this Section, we present the details of the linear stability analysis used in Fig.~(4) of the main text
to determine the lifetime of the subradiant states.

\subsection*{Determining the stability of the fixed points}
The fixed points $\psi_0$ ($\partial_t \psi_0=0$) of the GGPE (\ref{eq:GGPE}), including the subradiant states, obey the equation
\begin{equation}
    \left(-\frac{\hbar^2 \nabla^2}{2m} + H_\mathrm{mf}[\psi_0]\right)\psi_0 = 0\label{eq:steady-state_condition_psi_0}\,.
\end{equation}
In order to determine whether a fixed point is stable, one needs to perform a linear stability analysis. For a detailed discussion of this method, we refer the reader to textbooks about non-linear dynamics, such as Ref~\cite{Complex_and_adaptive_Systems_Gros_2015} 
and references therein.

The stability of $\psi_0$ can be determined by analyzing the dynamics of small fluctuations $\delta\psi(\bm{x},t) = \psi(\bm{x},t) - \psi_0(\bm{x})$ about the fixed point. As follows from the GGPE, the fluctuation obeys the exact equation of motion
\begin{equation}
    i\hbar \frac{\partial}{\partial t}\delta\psi(\bm{x},t) = \left(-\frac{\hbar^2 \nabla^2}{2m} + H_\mathrm{mf}[\psi_0 + \delta\psi]\right)\left(\psi_0(\bm{x}) + \delta\psi(\bm{x},t)\right)\label{eq:eom_delta_psi}\,.
\end{equation}

The fluctuations give rise to small variations of the observables' expectation values about their steady value $\langle \theta_\mu \theta_\nu \rangle_0 = \int\mathrm{d}\bm{x}~\psi_0^*(\bm{x})\theta_\mu(\bm{x})\theta_\nu(\bm{x})\psi_0(\bm{x})$ ~\cite{Jaeger:2019}, 
 where $\theta_\mu(\bm{x}) = \cos(\bm{k}_\mu \cdot \bm{x})$. We now use perturbation theory in first order in those variations, hereafter denoted by
\begin{equation}
    \langle \theta_\mu \theta_\nu \rangle_{\delta \psi} = \int \mathrm{d}\bm{x}~\theta_\mu(\bm{x})\theta_\nu(\bm{x})\left(\delta\psi^*(\bm{x},t)\psi_0(\bm{x}) + \psi_0^*(\bm{x})\delta\psi(\bm{x},t)\right)\label{eq:def_variation_observable}\,.
\end{equation}
The cavity amplitude $\alpha_1[\psi_0 + \delta\psi]$ to first order in the variations reads
\begin{equation}
    \alpha_1[\psi_0 + \delta\psi] = \alpha_1[\psi_0] + \delta \alpha_1\label{eq:expansion_alpha_1}\,,
\end{equation}
with
\begin{equation}
\begin{split}
    \delta \alpha_1 = & a_{1,11}[\psi_0]\langle \theta_1^2\rangle_{\delta\psi} + a_{1,22}[\psi_0]\langle \theta_2^2\rangle_{\delta\psi} + a_{1,p1}[\psi_0]\langle \theta_p\theta_1\rangle_{\delta\psi} + \\
    &+ a_{1,p2}[\psi_0]\langle \theta_p\theta_2\rangle_{\delta\psi} + a_{1,12}[\psi_0]\langle \theta_1\theta_2\rangle_{\delta\psi}\label{eq:expansion_alpha_1_first_order}\,.
\end{split}
\end{equation}
The expansion coefficients are given by
\begin{equation}
\begin{split}
    a_{1,11}[\psi_0] &=  \frac{U_1 N Z_{2,\mathrm{ss}}}{U_1U_2N^2 \langle \theta_1\theta_2 \rangle_0^2 - Z_{1,\mathrm{ss}}Z_{2,\mathrm{ss}}}\alpha_1[\psi_0]\,, \\
    a_{1,22}[\psi_0] &=  \frac{U_2 N\left(\sqrt{V_p U_1} N \langle \theta_p\theta_1 \rangle_0 + Z_{1,\mathrm{ss}}\alpha_1[\psi_0]\right)}{U_1U_2N^2 \langle \theta_1\theta_2 \rangle_0^2 - Z_{1,\mathrm{ss}}Z_{2,\mathrm{ss}}}\,, \\
    a_{1,p1}[\psi_0] &= \frac{\sqrt{V_p U_1} NZ_{2,\mathrm{ss}}}{U_1U_2N^2 \langle \theta_1\theta_2 \rangle_0^2 - Z_{1,\mathrm{ss}}Z_{2,\mathrm{ss}}}\,, \\
    a_{1,p2}[\psi_0] &= -\frac{U_2N\sqrt{V_p U_1} N\langle \theta_1\theta_2 \rangle_0}{U_1U_2N^2 \langle \theta_1\theta_2 \rangle_0^2 - Z_{1,\mathrm{ss}}Z_{2,\mathrm{ss}}}\,,\\
    a_{1,12}[\psi_0] &= -\frac{U_2 N\left(\sqrt{V_p U_1}N\langle \theta_p\theta_2 \rangle_0 + 2U_1 N \langle \theta_1\theta_2 \rangle_0\alpha_1[\psi_0]\right)}{U_1U_2N^2 \langle \theta_1\theta_2 \rangle_0^2 - Z_{1,\mathrm{ss}}Z_{2,\mathrm{ss}}}\,,\label{eq:coeffs_a}
\end{split}
\end{equation}

where $Z_{\nu,\mathrm{ss}} = U_\nu N \langle \theta_\nu^2 \rangle_0 - \Delta_\nu -i \kappa_\nu$. The expansion for the cavity $2$ amplitude $\alpha_2$ can be directly obtained from Eqs.~(\ref{eq:expansion_alpha_1}) - (\ref{eq:coeffs_a}) by interchanging the cavity labels $1 \leftrightarrow 2$.
Using these expressions in the mean-field Hamiltonian and keeping only the first-order terms, we find
\begin{equation}
    H_\mathrm{mf}[\psi_0 + \delta\psi] =
    H_\mathrm{mf}[\psi_0]+ \delta H_\mathrm{mf}[\psi_0]\label{eq:expansion_Hmf}\,,
\end{equation}
with 
 \begin{equation}
 \delta H_\mathrm{mf}[\psi_0]
    \approx \hbar\sum_{(\mu,\nu) \in \mathcal{S}}\sum_{(n,m) \in \mathcal{S}}h_{nm,\mu\nu}[\psi_0]\langle \theta_\mu \theta_\nu\rangle_{\delta\psi}\theta_n \theta_m\,.\label{eq:expansion_Hmf_first_order}
\end{equation}
Here, $\mathcal{S}=\{(1,1),(2,2),(p,1),(p,2),(1,2)\}$ is the set of all possible index tuples and
\begin{gather}
    h_{11,\mu\nu}[\psi_0] = 2U_1 \mathrm{Re}\left\{a_{1,\mu\nu}\left(\alpha_1[\psi_0]\right)^*\right\}\,,\quad h_{22,\mu\nu}[\psi_0] = 2U_2\mathrm{Re}\left\{a_{2,\mu\nu}\left(\alpha_2[\psi_0]\right)^*\right\}\,,\nonumber\\
    h_{p1,\mu\nu}[\psi_0] = 2\sqrt{V_p U_1}\mathrm{Re}\left\{a_{1,\mu\nu}\right\}\,, \quad h_{p2,\mu\nu}[\psi_0] = 2\sqrt{V_p U_2}\mathrm{Re}\left\{a_{2,\mu\nu}\right\}\,,\label{eq:coeffs_h}\\
    h_{12,\mu\nu}[\psi_0] = 2\sqrt{U_1 U_2}\mathrm{Re}\left\{\left(a_{1,\mu\nu}\right)^*\alpha_2[\psi_0]+\left(\alpha_1[\psi_0]\right)^*a_{2,\mu\nu}\right\}\,.\nonumber
\end{gather}
Plugging \eqref{eq:expansion_Hmf} in the equation of motion (\ref{eq:eom_delta_psi}) and accounting for the steady-state condition (\ref{eq:steady-state_condition_psi_0}), we obtain
\begin{equation}
    i\hbar \frac{\partial}{\partial t}\delta\psi(\bm{x},t) = \left(-\frac{\hbar^2 \nabla^2}{2m} + H_\mathrm{mf}[\psi_0]\right)\delta\psi(\bm{x},t) + \delta H_\mathrm{mf}[\psi_0]\psi_0(\bm{x})\label{app:eom:deltapsi}\,.
\end{equation}
Equation (3) in the main text corresponds to the Fourier transform of this equation. To analyze the spectrum of this dynamical equation, encoding the behaviour of the fluctuation $\delta\psi$, and thus the stability of the fixed point $\psi_0$, we follow the steps outlined in Ref~\cite{Jaeger:2019}. 
We first formally integrate \eqref{app:eom:deltapsi}, yielding
\begin{equation}
    \delta\psi(\bm{x},t) = e^{-\frac{i}{\hbar}H_0 t}\delta\psi(\bm{x},t=0) -\frac{i}{\hbar} \int_0^t\mathrm{d}\tau~e^{-\frac{i}{\hbar}H_0(t-\tau)}\delta H_\mathrm{mf}[\psi_0]\psi_0(\bm{x})\,,
\end{equation}
with $H_0 = -(\hbar^2 \nabla^2)/(2m) + H_\mathrm{mf}[\psi_0]$. Note that the quantity $\delta H_\mathrm{mf}[\psi_0]$ depends on the time $\tau$ via the variations $\langle \theta_\mu \theta_\nu\rangle_{\delta\psi}$ (see \eqref{eq:expansion_Hmf_first_order}). Applying the Laplace transform $L[f](s) = \int_0^\infty \mathrm{d}t e^{-st}f(t)$, $\quad s \in \mathbb{C}$, to the solution, while accounting for the convolution theorem of the Laplace transform~\cite{Laplace_transform_textbook}, 
one obtains
\begin{equation}
\begin{split}
    L[\delta\psi](s) = & \left(s +i H_0/\hbar\right)^{-1}\delta\psi(\bm{x},t=0) \\
    & -i \sum_{(\mu,\nu) \in \mathcal{S}}\sum_{(n,m) \in \mathcal{S}}L[\langle \theta_\mu \theta_\nu\rangle_{\delta\psi}](s)h_{nm,\mu\nu}[\psi_0]\left(s +i H_0/\hbar\right)^{-1}\theta_n \theta_m\psi_0(\bm{x})\,.
\end{split}
\end{equation}
Together with the Laplace transform of the complex conjugated fluctuation $\delta\psi^*$, we find an equation for the Laplace transformed variation $L[\langle \theta_\alpha \theta_\beta\rangle_{\delta\psi}](s)$:
\begin{equation}
    L[\langle \theta_\alpha \theta_\beta\rangle_{\delta\psi}](s) = b_{\alpha\beta}(s)-\sum_{(\mu,\nu) \in \mathcal{S}}\sum_{(n,m) \in \mathcal{S}}C_{\alpha\beta,nm}h_{nm,\mu\nu}[\psi_0]L[\langle \theta_\mu \theta_\nu\rangle_{\delta\psi}](s)\label{eq:Laplace_variation}\,,
\end{equation}
with
\begin{align}
    b_{\alpha\beta}(s) &= \int \mathrm{d}\bm{x}~ \theta_\alpha\theta_\beta\left(\psi_0(s-iH_0/\hbar)^{-1}\delta\psi^*(t=0) + \psi_0^*(s+iH_0/\hbar)^{-1}\delta\psi(t=0)\right)\,,\label{eq:def_b}\\
    C_{\alpha\beta,nm}(s) &= i\int\mathrm{d}\bm{x}~ \theta_\alpha\theta_\beta\left(\psi_0^*\left(s +i H_0/\hbar\right)^{-1}\theta_n \theta_m\psi_0 - \psi_0\left(s - i H_0/\hbar\right)^{-1}\theta_n \theta_m\psi_0^*\right)\,.\label{eq:def_C}
\end{align}
Note that we have omitted here for the sake of presentation the position argument $\bm{x}$ of the functions appearing in the integrals.  Equation (\ref{eq:Laplace_variation}) describes a set of five coupled equations for the Laplace transformed variations $L[\langle \theta_\alpha \theta_\beta\rangle_{\delta\psi}]$, with $(\alpha,\beta) \in \mathcal{S}$. They can be conveniently combined in a single matrix-vector equation $(1 + C(s)h)\bm{Y}(s) = \bm{b}(s)$, where $\bm{Y}(s)=(L[\langle \hat{\theta}_1^2\rangle_{\delta\psi}](s),L[\langle \hat{\theta}_2^2\rangle_{\delta\psi}](s),L[\langle \hat{\theta}_p\hat{\theta}_1\rangle_{\delta\psi}](s),L[\langle \hat{\theta}_p\hat{\theta}_2\rangle_{\delta\psi}](s),L[\langle \hat{\theta}_1 \hat{\theta}_2\rangle_{\delta\psi}](s))^T$ and the entries 
of the vector $\bm{b}$, and the matrices $C$ and $h$ are given in Eqs. (\ref{eq:coeffs_h}), (\ref{eq:def_b}), and (\ref{eq:def_C}). The stability of the fixed point can be determined by analyzing the solutions $s$ of the dispersion relation
\begin{equation}
    \mathrm{det}(1+C(s)h)=0\,,\label{eq:dispersion_relation}
\end{equation}
as discussed in Ref~\cite{Jaeger:2019}
and in the following section.

\subsection*{Extracting the subradiant state's lifetime}
The subradiant state represents a fixed point of the GGPE (\ref{eq:GGPE}), thus fulfilling the steady-state condition (\ref{eq:steady-state_condition_psi_0}). As such, we can employ the linear stability analysis of the preceding section to determine its stability for given parameters. As we will discuss in the following, this allows us to predict its lifetime in the presence of atom losses.

In Fig.~4 of the main text, we study quenches from cavity $2$ to $1$. Here, the subradiant state of the system (after performing the quench) is characterized by a spatial configuration that supports the scattering of photons into cavity $2$ but suppresses scattering into cavity $1$. The corresponding steady state $\psi_0$ can be computed by solving the steady-state condition (\ref{eq:steady-state_condition_psi_0}) for the cavity detunings after the quench, $(\Delta_1,\Delta_2) = (0,\delta)$, and for vanishing coupling to cavity $1$, $g_1 = 0$. We verify that this state obeys our definition of the subradiant state, that is, $\langle\theta_p\theta_1\rangle_0 = \langle\theta_1\theta_2\rangle_0 = 0$. Hence, the spatial configuration prohibits the occupation of this cavity, such that $\alpha_1[\psi_0] = 0$, and the state $\psi_0$ remains a steady state of the dynamical equation, even in the presence of the coupling to cavity $1$, $g_1 \neq 0$. By numerically computing the roots of the dispersion relation (\ref{eq:dispersion_relation}) for the state $\psi_0$ and for $g_1 \neq 0$, we determine its stability. A root with positive real part $\gamma > 0$ indicates that the coupling to cavity $1$ renders the subradiant state unstable, otherwise it remains stable. Repeating this calculation for different quench strengths $\delta$ and atom numbers $N$, we construct the stability diagram reported in Fig.~4\textbf{F} in the main text.

As visible from this stability diagram, we can identify for each considered quench strength $\delta$ a critical atom number $N_c(\delta)$ below which the subradiant state becomes unstable ($\gamma > 0$). Suppose that the atom number after the quench is $N_0$ and takes values between $2.43\times 10^5$ and $2.97 \times 10^5$, corresponding to the typical range of atom numbers in the experiment. Then the subradiant state represents a stable configuration if $N_0 > N_c$. Even if initially fulfilled, atom losses eventually decrease the atom number over time, $N(t) \leq N_0$, and the subradiant state becomes unstable when $N(t^*) = N_c$. Thus, the atom loss brings the system from the stable region ($\gamma \leq 0$) to the unstable region ($\gamma > 0$) in the stability diagram, as indicated by the green arrow in Fig.~4\textbf{F}. 
The delay times shown in Fig.~4\textbf{E} 
represent essentially the length of the time interval it takes the atom losses to decrease the initial atom number below the critical value. Imposing as above an exponential decay of the atom number, $N(t) = N_0e^{-\Gamma t}$, we can predict the delay time $t^*$ by solving the condition $N_0e^{-\Gamma t^*} = N_c$, yielding
\begin{equation}
    t^*(\delta) = -\frac{1}{\Gamma}\ln\left(\frac{N_c(\delta)}{N_0}\right)\,.
\end{equation}
It is important to note that this equation only allows us to make viable predictions about the delay time if the dynamical system state $\psi(t)$ remains close to the predicted steady subradiant state $\psi_0$ after the quench. In particular, as the latter is a function of the atom number, $\psi_0 = \psi_0(N)$, the subradiant state evolves over time when atom loss is present. As the atom loss occurs here on a much longer timescale than the one associated with the long-range interaction, the system - if initially trapped in the subradiant state - is able to follow adiabatically the evolving subradiant state.

\section*{Experimental Details}
\label{sec:Experimental Details}
\subsection*{Preparation of the Bose Einstein condensate (BEC) and balancing the two atom-cavity couplings}

Our experimental protocol starts with the preparation of a $^{87}$Rb Bose-Einstein condensate (BEC) of $N=2.7(3)\times 10^5$ and $T=102(10)\,\mathrm{nK}$ via optical evaporation. An optical dipole trap with trapping frequencies $[ \omega_\textrm{x}, \omega_\textrm{y}, \omega_\textrm{z}]=2\pi \times [106(1), 71.2(3), 223(3)]\,\mathrm{Hz}$ positions the BEC at the crossing point of the two cavity modes. 

The atom-cavity coupling is proportional to the overlap of the cavity mode with the atomic cloud, and can be measured via the dispersive shift of the cavity resonance. Moving the trap to different locations and measuring the dispersive shift as a function of the trap position allows us to map out the effective coupling strengths for both resonators. Since the cavity mode centers differ by 12~$\mu$m in height, where the mode waist radius is 50~$\mu$m, we can adjust the trap position such that the coupling to both resonators is balanced. Slight misalignments can however lead to a systematic shift of the phase boundary between the two ordered phases. The data in Figure 2 has therefore been shifted by 0.1~MHz to compensate for this systematic error.

A magnetic offset field of $\sim25$~G is applied to avoid spin-dependent effects on self-organization as e.g. superradiance to the orthogonally polarized cavity mode that is detuned due to the birefringence of our cavity as we studied in~\cite{morales2019two}.

\subsection*{Cavities and phase diagrams}
The cavity modes are tilted at an angle of $60^{\circ}$ and have distinct decay rates $[ \kappa_\textrm{1}, \kappa_\textrm{2}]=2\pi \times [147(4), 800(11)]\,\mathrm{kHz}$. The maximum single-atom vacuum Rabi frequencies of $[ g_\textrm{1}, g_\textrm{2}]=2\pi \times [1.95(1), 1.77(1)]\,\mathrm{MHz}$ are comparable for both cavities.
At a $60^{\circ}$ angle relative to each cavity, a retroreflected pump beam creates a standing wave lattice. We refer to this beam as the transverse pump. Its frequency $\omega_\textrm{p}$ is blue detuned by $\Delta_\textrm{a}=\omega_\textrm{p}-\omega_\textrm{a}=2\pi \times 69.8(1)\,\mathrm{GHz}$ with respect to $\omega_\textrm{a}$, the $\text{D}_2$-line of $^{87}$Rb. We further introduce the relative detunings $\Delta_\nu=\omega_\textrm{p}-\omega_\nu^c$ (with $\nu\in \{1,2\}$) of the cavity resonance frequencies $\omega_\nu^c$ and the transverse pump. These detunings between the frequency of the transverse beams and the cavities are tuneable within tens of MHz via an electro-optical modulator. The lengths of the cavities are stabilized using low amplitude, far-detuned additional laser fields at $830 \, \text{nm}$,  allowing a constant feedback on the cavity lengths while having a negligible effect on the atomic cloud.

During each experimental sequence, we linearly increase the transverse pumping lattice depth $V_\textrm{p}$ in $10 \,\mathrm{ms}$ from $0$ to a maximum of $10.8 \,\omega_\textrm{r}$, in units of recoil frequency, $\omega_\textrm{r}= 2\pi \times 3.77\,\mathrm{kHz}$. Meanwhile, we collect the leaking photons from each cavity and use the photon detection rates to deduce the intracavity photon numbers. Beyond a critical pump strength and for certain cavity detunings, the intracavity field is populated as the atoms self-organise into minima of the resulting lattice potential. For each cavity we repeat the transverse pumping ramp for different cavity detunings and obtain the single cavity phase diagrams as shown in Figure~S1 a, d. The respective other cavity is not influencing this measurement since we put it to a sufficiently large detuning. The atom-cavity couplings are well balanced for both cavities, resulting in similar phase boundaries for self-organisation. We further extend the individual phase diagrams to two cavities simultaneously close to resonance. By increasing the transverse pump power, we study the self-organisation for different detunings of cavity 1 and 2 at a fixed transverse pump lattice strength of 5.1 $\omega_\textrm{r}$. We plot the photon numbers in cavity 1 (2) in Figure S1 b (e) for the given pair of cavity detunings. If the cavity field in cavity 1 (2) is larger than a defined photon number threshold, $n_1^\mathrm{th}$ ($n_2^\mathrm{th}$), we characterise the system as self-organised in cavity 1 (2) and plot a blue (red) square in the binarized Figure~S1 c for the given pair of cavity detunings. A white square indicates photon levels in both cavities below the thresholds, and a grey square indicates photon levels in both cavities above the thresholds. Self-organisation occurs in the cavity closer to resonance. To further highlight this dependence, we switch to a rotated coordinate system and define the relative detuning $\Delta_\textrm{1}-\Delta_\textrm{2}$ and the mean detuning ($\Delta_\textrm{1}+\Delta_\textrm{2}$)/2 as the x and y axes, see Figure~S1 f.

\subsection*{Heterodyne measurement}

The amplitudes of the intracavity fields are recorded using balanced optical heterodyne detectors. For each cavity, a local oscillator (LO) laser field is combined at a beam splitter with the light field leaking from a cavity mirror and directed together onto the two photodiodes of a balanced photodetector. The signal is then mixed down to a moderate radio frequency of $300\,\text{kHz}$ using a homemade IQ filter device. The LO is generated by shifting the frequency of the same laser source as the transverse pump by $50\,\text{MHz}$ with an acousto-optic modulator. To compensate for phase drifts introduced by sending the light fields through separate optical fibres, we lock their relative phase after passing through the fibres. For each cavity signal, the $300\,\text{kHz}$ data is recorded using a computer connected oscilloscope (PicoScope 5444b). The signal is then processed by software which extracts the photon level and applies a low pass filter using a binning window of $1\times10^{-4}$ s. All associated radio frequency signal sources are phase-locked to a $10\,\text{MHz}$ GPS frequency standard which has a fractional stability better than $10^{-12}$ to minimise the technical phase noise in the heterodyne detection system. Further technical details and design considerations are described in detail in~\cite{li2021first}. To calibrate the amplitude of the intracavity light field, we perform Raman-Nath diffraction analogous to the transverse pump calibration by applying a short coherent on-axis probe to the cavity.

\subsection*{Imaging of the atomic cloud}

We use resonance absorption imaging of the atomic cloud in addition to the detection of the light field leaking from the cavities. From absorption images taken after $20 \, \text{ms}$ of ballistic expansion of the atomic cloud, the temperature, number of atoms and momentum state populations are extracted.

\subsection*{Atomic loss rate}
We measure the loss rate of the system by preparing the cloud in the crossed beam dipole trap and ramping up the transverse pump lattice. We let the system evolve for a certain hold time before we switch off all potentials and measure the remaining atom number after ballistic expansion. Typical decay curves are shown in Figure~\ref{fig:atomloss} for different parameters.

\subsection*{Measurement protocol of the subradiant states}
In the section "Formation of Subradiant States" in the main text, we present a measurement protocol which extracts a hysteresis region of subradiant states. For certain cavity 1 and 2 detunings, we linearly ramp up the transverse pump power in $5 \,\mathrm{ms}$ such that the atoms self-organize into the initial cavity. Next, over $25 \,\mathrm{ms}$, the cavity detunings are swept across the phase boundary. The sweep is symmetric with regards to the phase boundary: one of the ramps linearly changes $\Delta_\textrm{1,2}$ from $[0.5, -0.5]\,2\pi \times \mathrm{MHz}$ to $[-0.5, 0.5]\,2\pi \times \mathrm{MHz}$ and yields $\Delta_\textrm{1}-\Delta_\textrm{2}$ as a unique parameter characterizing this sweep. At the sweep half-time of $12.5 \,\mathrm{ms}$, the cavity detunings are degenerate $\Delta_\textrm{1}-\Delta_\textrm{2}=0$. We extract the switching time from traces as shown in Fig.~3\textbf{A},
which corresponds to a unique value of $\Delta_\textrm{1}-\Delta_\textrm{2}$. We repeat this measurement for different values of $(\Delta_\textrm{1}+\Delta_\textrm{2})/2$ and show the results in Fig.~3\textbf{B}.

In the section "Lifetime of Subradiant States" in the main text, we characterize the lifetimes of the subradiant system by quenching across the phase boundary between the two cavities and determining the time after which the system switches between the two cavities (see Fig.~4
). At the start of each quench measurement, a $10 \,\mathrm{ms}$ long transverse pump ramp is used to prepare a self-organized state in one of the cavities. The detunings of the cavities determine which cavity gets populated. When quenching from cavity 2 to 1, cavity 2 is initially resonant with respect to the transverse pump, so $\Delta_\textrm{2}=0$ and cavity 1 is detuned to $\Delta_\textrm{1}=[-0.6, -0.1]\,2\pi \times \mathrm{MHz}$. After the transverse pump ramp, we hold the final pumping strength and cavity 1 is quenched onto resonance $\Delta_\textrm{1}=0$ and cavity 2 away from resonance by $\Delta_\textrm{2}=[-0.6, -0.1]\,2\pi \times \mathrm{MHz}$. Typical photon traces for both cavities are shown in Fig.~4\textbf{A,~C}. 
Following a buildup of the photon level in cavity 2, the system stays self-organized in cavity 2 beyond the quench. It remains subradiant until a characteristic delay time, when the self-organization jumps from the second to the first cavity. We extract this delay time from numerically fitting a delta function in cavity 2 to the decay and in cavity 1 to the onset of the photon field. We find that both times lay within $0.1\,\mathrm{ms}$. We nevertheless take the average of both as the delay time plotted in Fig.~4\textbf{E}. 

\begin{figure}[]
\includegraphics[width=\textwidth]{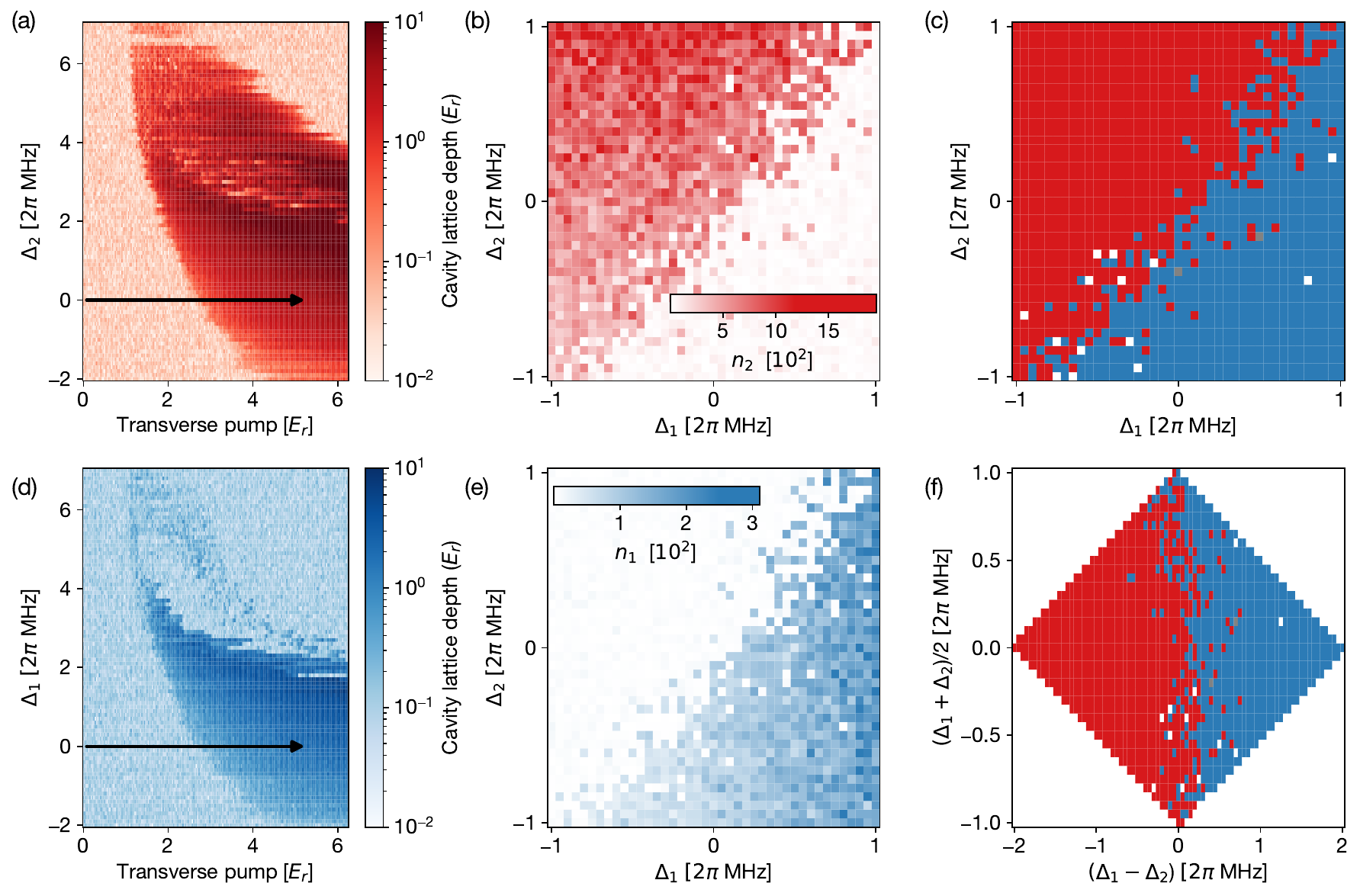}
\caption{\textbf{Measuring Phase Diagrams.} \textbf{(a,d)} Single cavity phase diagrams for cavity 2 (1). We linearly ramp up the transverse pump to $6.2~ \text{E}_r$ and record the cavity lattice depth for different detunings $\Delta_2$ ($\Delta_1$). For these measurements, cavity 1 (2) is detuned far away from resonance and does not influence the self-organisation of cavity 2 (1). Black arrows highlight transverse pump ramps, where we stop the ramp at $5.1~ \text{E}_r$ and collect data for the subsequent panels. \textbf{(b,e)} Measured intracavity photon number $\textrm{n}_\textrm{1(2)}$ after increasing $\hbar V_p$ linearly from 0 to $5.1~ \text{E}_r$ at fixed detunings $\Delta_\textrm{1(2)}$ \textbf{(c)} Combined and binarized phase diagram derived from panels \textbf{(b,e)} by defining thresholds of $(n_1^\textrm{th},n_2^\textrm{th}) = (30, 200)$ photons for identifying superradiance in a respective cavity. The steady-state photon numbers, and thus defined threshold values, differ due to the different cavity decay rates. Blue (red) pixels indicate superradiance in cavity 1 (2), white pixels indicate both cavity populations below threshold, grey pixels indicate coexistence where both cavity populations are above the respective threshold. \textbf{(f)} Same data as in panel \textbf{(c)}, but in rotated coordinates of relative cavity detunings $\Delta_\textrm{1}-\Delta_\textrm{2}$ and mean cavity detunings $(\Delta_\textrm{1}+\Delta_\textrm{2})/2$ as used in the main text. White spaces outside the square indicate no data.}
\label{fig:extractingphasdiagrams}
\end{figure}

\begin{figure}[]
\includegraphics[width=\textwidth]{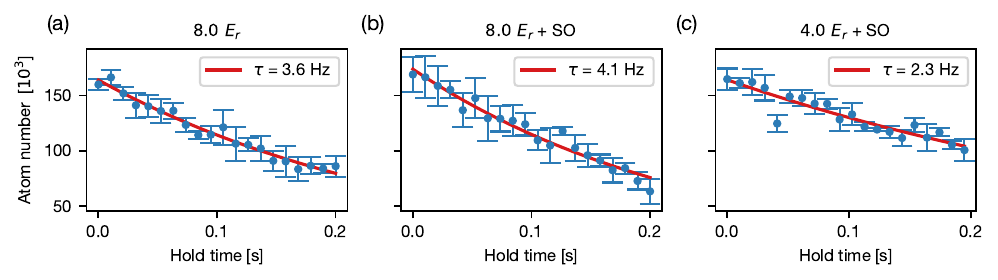}
\caption{\textbf{Atom loss.} The transverse pump drive is linearly ramped up in $10~ \text{ms}$ from $0.0$ to $8.0~ \text{E}_r$ for \textbf{(a,b)} and from $0$ to $4.0~ \text{E}_r$ for \textbf{(c)}. After variable hold time, the atom numbers are extracted from time-of-flight images of the cloud. The cavity 1 detuning in \textbf{(a)} is detuned away from resonance and atoms do not self-organise. For \textbf{(b,c)}, the cavity 1 detuning is set to resonance such that atoms self-order during the transverse pump ramp. In all measurements cavity 2 is far detuned from resonance. This figure estimates atom loss rates of up to $4~ \text{Hz}$ present in our experiment, which have been taken into account in the numerical models.}
\label{fig:atomloss}
\end{figure}

\end{appendix}

\end{document}